\documentclass[traditabstract,referee]{aa}
\usepackage{txfonts}
\usepackage{graphicx}
\usepackage{natbib}

\def\kms{\hbox{km\,s$^{-1}$}}
\def\pcc{\hbox{cm$^{-3}$}}
\def\Bs{\hbox{$B_{\rm s}$}}
\def\Bl{\hbox{$B_{ \ell}$}}
\def\deg{\hbox{$^\circ$}}
\def\omeq{\hbox{$\Omega_{\rm eq}$}}
\def\dom{\hbox{$d\Omega$}}
\def\msun{\hbox{${\rm M}_{\odot}$}}

\def\mstar{\hbox{$M_{\star}$}}
\def\rstar{\hbox{$R_{\star}$}}
\def\mphi{\hbox{$M_{\Phi}$}}

\begin{document}

\title{Magnetic fields of non-degenerate stars}
\author{J.-F. Donati\inst{1} \and J.D.\ Landstreet\inst{2}}
\institute{CNRS/Universit\'e de Toulouse, France \and University of Western Ontario, Canada}
\abstract{Magnetic fields are present in a wide variety of stars throughout the HR diagram 
and play a role at basically all evolutionary stages, from very-low-mass dwarfs to very 
massive stars, and from young star-forming molecular clouds and protostellar accretion discs 
to evolved giants/supergiants and magnetic white dwarfs/neutron stars.  
These fields range from a few $\mu$~G (e.g., in molecular clouds) to TG and more (e.g., in 
magnetic neutron stars);  in non-degenerate stars in particular, they feature large-scale 
topologies varying from simple nearly-axisymmetric dipoles to complex non-axsymmetric structures, 
and from mainly poloidal to mainly toroidal topology.  
After recalling the main techniques of detecting and modelling stellar magnetic fields, 
we review the existing properties of magnetic fields reported in cool, hot and young 
non-degenerate stars and protostars, and discuss our understanding of the origin 
of these fields and their impact on the birth and life of stars.  }
\keywords{stars: magnetic fields;  stars: formation;  stars: late-type;  stars: early-type;  
          stars: rotation;  stars: activity;  stars: abundances;  stars: mass-loss;  
          techniques: spectroscopy;  techniques: polarimetry}
\maketitle 

\section{Introduction} 

Magnetic fields are known to be present in a wide variety of stars, from 
very low-mass M dwarfs to super-massive O stars.  They play a role at 
basically all evolutionary stages, from collapsing molecular clouds and very 
young protostars to supernovae, degenerate white dwarfs and neutron stars 
\citep[e.g.,][]{Mestel99}.  
Magnetic fields are found to influence significantly a number 
of physical processes operating within and in the immediate vicinity of 
stars, such as accretion, diffusion, mass loss, turbulence and 
fundamental quantities such as mass, rotation rate and 
chemical composition.  The aim of this paper is to summarise the main observational 
results in this field of research, especially those obtained since the previous 
review of \citet{Landstreet92}, and to outline our current
understanding of how magnetic fields impact the lives of stars 
of various masses.  

The first detection of a magnetic field in a star, our Sun, was obtained one 
century ago by \citep{Hale08} who observed and correctly interpreted the magnetic 
polarisation of spectral lines in sunspots and attributed it to magnetic fields of 
nearly 3~kG.  This was the first astrophysical 
application of the Zeeman effect, discovered by Zeeman only 12~yr before 
the pioneering work of Hale.  Searching for magnetic stars other than the Sun 
to investigate how ``normal'' the Sun is, Babcock discovered in 1947 the simple 
large-scale fields of chemically peculiar stars \citep{Babcock47} and quickly 
realised that these magnetic fields are in fact fairly different in nature than 
that of the Sun.  About twenty years later, pulsars were detected for the first 
time through their radio emission \citep{Hewish68} and identified with rotating 
neutron stars \citep{Gold68} with tremendously intense fields of typically TG 
($10^{12}$~G) strengths, i.e., about one billion times stronger than those found 
at the heart of sunspots;  shortly afterwards, magnetic fields of several hundreds 
of MG were reported to exist at the surfaces of some white dwarfs \citep{Kemp70}.  
Magnetic fields on cool solar-type stars other than the Sun were suspected for a 
long time;  the solar analogy indeed suggests that the Sun-like activity observed 
on basically all low-mass stars is due to magnetic fields.  After a search of 
several decades, the first direct detections were  only obtained in 1980 
\citep{Robinson80}, more than 30~yr after Babcock's initial discovery.  

Since these initial results, our understanding of stellar magnetic fields has 
tremendously improved, thanks mostly to the improvement of instrumental 
performances and to the sophistication of numerical simulations.  
We know for instance that the magnetic field of the Sun is highly complex, 
giving rise not only to sunspot \citep[e.g.,][]{Solanki03} but also to a 
complex set of magnetic features of smaller sizes such as ephemeral regions, 
bright points or inter-network fields \citep[e.g.,][]{Stenflo89, Lites08}.  
Most cool stars also host magnetic fields, whose large- and medium-scale 
topologies can be revealed through tomographic techniques 
\citep[e.g.,][]{Landstreet92, Donati08e}.  Magnetic fields of low-mass stars power
a wide variety of energetic phenomena (referred to as activity) such as flares, 
prominences, coronae and winds;  the fields themselves most likely result from 
the interaction of convection and rotation (the so-called dynamo processes) and 
are highly variable in nature on timescales ranging from minutes (e.g., flares) 
to years (e.g., activity cycles).  They are responsible for slowing down the 
rotation of most cool stars once they reached the main sequence, on a timescale 
that depends mostly on the stellar mass.  

In contrast to low-mass stars which more or less all show magnetic fields and 
activity, intermediate- and high-mass stars are mostly non-magnetic with only a 
small minority (a few \% depending on spectral type) showing detectable magnetic 
fields \citep[e.g.,][]{Landstreet92}.  Moreover, these fields are fairly different 
in nature;  in particular, they often feature a fairly simple large-scale topology 
and show no intrinsic variability (other that the usual rotational modulation) 
even on timescales of decades.  Magnetic A and late B stars are all chemically 
peculiar stars;  the last decade has revealed that a small fraction of early B 
and O stars are also magnetic despite having nearly normal abundances.  
Magnetic fields in hot stars likely have a different origin, and their impact 
on evolution, although clear (e.g., abundance inhomogeneities likely result 
from their being magnetic) is still mostly a matter of speculation.  

The impact of magnetic fields is actually greatest during the formation stage, 
when the magnetic energies of molecular clouds are comparable to their 
gravitational energies \citep{Crutcher99}.  Once they become visible at optical 
wavelengths (i.e., at an age of about 1~Myr for low-mass stars), low-mass 
protostars host multi-kG magnetic fields at their surfaces through which they 
connect to (and accrete matter from) the inner rim of their accretion disc 
\citep[e.g.,][]{Bouvier07}.  Observations also reveal that protostellar 
accretion discs often exhibit broad outflows and highly-collimated jets that 
models can only reproduce by invoking magnetic fields redirecting a fraction 
of the accreted material into the outflows/jets through a propeller-type 
mechanism powered by magneto-centrifugal forces \citep[e.g.,][]{Pudritz07}.  
While it is not yet fully clear how massive stars are formed, 
observations of hot magnetic protostars suggest that they rotate 
significantly more slowly than normal and thus that magnetic fields have altered 
the cloud collapse and contraction stage towards the main sequence.  

At the other end of the evolution, white dwarfs and neutron stars can exhibit 
extraordinarily intense magnetic fields, with magnetars hosting the strongest 
(PG, i.e., $10^{15}$~G) fields known in the universe.  
Their magnetic fluxes, however, are similar to those of magnetic hot main sequence 
stars, suggesting empirically that most of the magnetic flux is conserved during 
stellar evolution off the main sequence and towards the very last stages.  
A large fraction of the 
white dwarfs with strong magnetic fields are found in close binary systems (called 
cataclysmic variables) where the companion is overflowing its Roche lobe and 
transferring mass to the white dwarf through magnetic funnels.  
Numerous studies describing advances in the field have been 
published recently \citep[e.g.,][]{Ferrario07};  however, for lack of space, 
we do not include here a description of the magnetic fields of 
degenerate stars, despite their obvious interest in understanding the fate 
of main sequence stars and the role of magnetic fields in evolutionary stages 
such as supernovae and planetary nebulae.  

This paper mostly focusses on magnetic field observations of non-degenerate stars 
of all masses.  We start by describing the methods by which magnetic fields 
in stars are detected and measured (in Secs.~\ref{sec:atom} and \ref{sec:inst});  
we then detail our current understanding of how magnetic fields are produced 
and affect the life of stars of various masses (in Secs.~\ref{sec:lowm} and \ref{sec:higm}) 
and in particular their formation (Sec.~\ref{sec:prot}).  While regularly referring 
to some of the basic magnetic properties of the Sun, we do not review all aspects of 
the solar magnetism, a field of research so wide and documented that it requires a 
complete review of its own.  For a more theoretical description of magnetic stars, 
readers are referred to the reference book of \citet{Mestel99} and to the forthcoming 
new edition.  

\section{Atoms and molecules in a magnetic field}
\label{sec:atom}

We briefly describe here the methods by which magnetic fields can be 
directly detected at the surface of non-degenerate stars\footnote{Indirect 
methods and proxies such as emission in various lines (e.g., Balmer lines, 
Ca H\&K and infrared triplet lines) or broad bands (UV, X-ray, radio) 
suggesting the presence of magnetic fields will be mentioned later, along 
with the discussion of the stellar class to which they mostly relate.}.    
Such detections/measurements are made possible by the effect of magnetic 
fields on atoms and molecules in stellar atmospheres, reflecting changes 
in the structure and energies of atomic/molecular energy levels and thus 
in the profiles and polarisation properties of stellar spectral lines. 

The Zeeman effect is the most well known of these effects and is used to 
diagnose all types of fields, from very weak (for instance the $\mu$G fields 
of molecular clouds) to very strong (the MG to GG fields of white 
dwarfs).  The vast majority of measurements and results reported in 
the literature has been obtained through Zeeman spectroscopy and 
spectropolarimetry.  
The Hanle effect is another example, more adapted to very weak tangled 
fields for which the Zeeman effect fails;  since it has never allowed yet 
to diagnose fields in stars other than the Sun, we will not discuss it here.  
Both effects apply on both atoms and molecules, and show different behaviour 
depending on how strong the field is.  
Only the main points are summarised here;  for more details, readers are 
referred to the comprehensive description of \citet{Mathys89} and to the 
reference textbook of \citet{Landi04}.  

Consider 
an isolated atom placed in a vector magnetic field {\bf B};  the modifications 
produced on the atom can be described by adding to the unperturbed Hamiltonian 
$H_0$ of the atomic system an additional term $H_B$ called the magnetic Hamiltonian 
and given by:  
\begin{equation}
\label{eq:ham1}
H_B = \frac{e h}{4 \pi m c} ({\bf L} + 2 {\bf S}) \cdot {\bf B} +
      \frac{e^2}{8 m c^2} ( {\bf B} \times {\bf r} )^2 
\end{equation}
where $m$ and $e$ are the electron mass and charge, $c$ is the speed of light, 
$h$ is the Planck constant, {\bf L}, {\bf S} and {\bf r} are the total orbital 
angular momentum, spin and position operators of the electronic cloud.  

The second (quadratic) term of Eq.~\ref{eq:ham1} is called the diamagnetic term;  
its importance is very limited in practice in stars other than white dwarfs and 
neutron stars that we do not discuss here.  In this context, $H_B$ simplifies to 
the linear term:  
\begin{equation}
\label{eq:ham2}
H_B = \mu_0 ({\bf L} + 2 {\bf S}) \cdot {\bf B} 
\end{equation}
where $\mu_0$ is the so-called Bohr magneton.  If the magnetic field is weak enough 
(typically smaller than 1~MG) to keep the magnetic energy smaller than the energy 
intervals relative to the unperturbed Hamiltonian $H_0$, the effect of $H_B$ can be 
computed by the perturbation theory.  

The well-known result is that each degenerate energy level 
of $H_0$ splits into $2J+1$ sublevels with energy shifts given by:  
\begin{equation}
\label{eq:enl1}
\Delta E = \mu_0 g B M \hspace{1cm} (M=-J,-J+1, ..., J) 
\end{equation}
where $g$ is a dimensionless factor (usually lying between 0 and 3) 
called the Land\'e factor of the atomic level.  Assuming the atomic 
levels are described by the Russell-Saunders (or $L-S$) coupling scheme (a good 
approximation for light atoms), $g$ can be written as;  
\begin{equation}
\label{eq:lan}
g_{LS} = 1 + \frac{ J(J + 1) + S(S+1) - L(L+1)}{2J(J+1)}.
\end{equation}

Transitions between a level $E_i$ and another level $E_f$ of Land\'e 
factors $g_i$ and $g_f$ are characterized by a single energy $E_f-E_i$ in 
the absence of a magnetic field.  When a field is applied, the spectral line 
splits into closely spaced components with energies shifted from 
the rest energy by: 
\begin{equation}
\label{eq:enl2}
\Delta E = (g_f M_f - g_i M_i) \mu_0 B  
           = (\Delta g M_f + g_i \Delta M) \mu_0 B  
\end{equation}
where $\Delta g = g_f - g_i$ and $\Delta M = M_f-M_i$.  

Dipole transitions between 
the levels obey the selection rule $\Delta M=0$, $-1$ or $+1$ and the resulting 
spectral lines form three corresponding groups.  The lines due to transitions 
$\Delta M = 0$ ($\pi$ components) are distributed symmetrically around the 
unsplit line formed in the absence of a field.  The two groups of lines formed 
by transitions with $\Delta M=\pm1$ ($\sigma$ components) are shifted symmetrically 
about the unsplit line, with transitions of $\Delta M=1$ on one side and transitions 
of $\Delta M=-1$ on the other side.  In general, $\pi$ and $\sigma$ groups have 
several components;  when they all overlap within each group (e.g., when $J_f=0$, 
$J_i=0$ or $g_f=g_i$), the transition is called a Zeeman triplet.  Some lines 
show no Zeeman splitting (e.g., $g_i=0$ and $J_f=0$) and are called 
magnetic null lines.  

The average wavelength displacement $\Delta \lambda_B$ (in pm, i.e., 0.001~nm) of a $\sigma$ 
component from its zero field wavelength $\lambda$ (in $\mu$m) for a magnetic field 
B (in kG) is given by:  
\begin{equation}
\label{eq:lam}
\Delta \lambda_B = 4.67 \lambda_0^2\ \bar{g} B  
\end{equation}
where $\bar{g}$ is called the effective Land\'e factor measuring the average 
magnetic sensitivity of the line;  in practice, values of $\bar{g}$ are typically 
about 1.2 but can range from 0 up to 3.  The actual size of the line splitting 
is about 1.4~pm (or 0.84~\kms) for a transition at 500~nm with $\bar{g}=1.2$ in 
a 1~kG field;  going to a wavelength of 2~$\mu$m in the near infrared (nIR), the 
splitting rises to about 22.4~pm (3.36~\kms) for the same field strength and 
Land\'e factor.  

If the magnetic energy is comparable to the energy intervals of fine structure splitting of $H_0$ 
(i.e., at field strengths larger than 100~kG), 
the perturbation theory is no longer applicable;  the Zeeman effect is entering 
a non-linear regime called the Paschen-Back regime.  The Paschen-Back regime has 
few astronomical applications in non-degenerate stars.  The fields found in
main sequence stars (generally of the order of 1~kG, very rarely as large as 30~kG) 
are not large enough to push most lines into the Paschen-Back regime.

Diatomic molecules (e.g., TiO, CH, C$_2$, OH, CN, MgH, CaH or FeH) are also 
sensitive to magnetic fields, generating Zeeman splitting and polarisation of spectral  
lines in a way similar to atoms;  they can also be used to detect and measure magnetic 
fields in stars cool enough to show molecular lines in their spectra.  
The study of the molecular Zeeman effect has, in contrast to the 
atomic effect, been largely neglected until recently \citep{Schadee78};  
efforts from different groups, both on the theoretical and experimental sides, have 
been carried out over the last decade \citep[e.g.,][]{Berdyugina02, Asensio06} to 
provide observers with updated tools for modelling unpolarized and polarized spectra 
of magnetic stars.  

In practice, the Zeeman effect in stellar spectra can be detected both through the
magnetic splitting or broadening of unpolarized spectral lines when the field 
strength is strong enough - stronger than 1~kG typically.  
It can also be detected as polarisation signals in spectral lines,  
even for fields as weak as a few G in the optical domain and a few $\mu$G at 
radio wavelengths;  however, polarisation being sensitive to the 
vector properties of the field, the Zeeman polarisation is basically insensitive to 
very tangled weak fields, in which regions of opposite polarities mutually 
cancel out their respective contributions.  

\section{Detecting, measuring \& modelling stellar magnetic fields}
\label{sec:inst}

Several techniques are currently used to detect, estimate and model 
magnetic fields at the surfaces of non-degenerate stars, reflecting 
mostly the different types of instrument that can be used for this 
purpose.  
While some methods use high-resolution spectroscopy to study the 
detailed shape of line profiles, others attempt to measure the 
polarisation of spectral lines (through either photopolarimetry or 
spectropolarimetry) that magnetic fields produce through the Zeeman 
effect.  
We describe below the various options and mention the typical 
magnetic diagnostics that they provide.  

\subsection{High-resolution spectroscopy}

The most direct and easily interpreted means of detecting magnetic fields 
in stars is by observing the Zeeman splitting of spectral lines.  In this 
respect, the most natural instrument is a high-resolution spectrometer 
with a resolving power of at least 50,000 (i.e., 6~\kms) and possibly as high 
as 100,000 (i.e., 3~\kms).  From the separation of the $\pi$ and $\sigma$ 
components of a line with known Land\'e factor $\bar g$, the intensity 
of the magnetic field averaged over the visible hemisphere of the star 
\Bs\ can be measured (see Eq.~\ref{eq:lam}).  However, in main sequence 
stars, the detection of Zeeman splitting is not easy;  the broadening to 
be measured competes with other sources of line broadening, the 
most important of which is usually rotation.  Even for the slowest rotators, 
turbulent broadening (e.g., in low-mass stars) can easily produce line widths 
of several \kms, implying that the splitting can only be detected for field 
strengths larger than typically 5~kG at optical wavelengths.  
In practice, Zeeman splitting of optical spectral lines is mostly observed 
in magnetic chemically peculiar stars with very slow rotation and 
negligible turbulent broadening, where surface fields of a few kG are 
routinely detected \citep[e.g.,][]{Mathys97}.  

In stars with weaker fields, the magnetic 
intensity may still be estimated through the broadening of magnetically 
sensitive spectral lines.  For a highly-magnetic ${\bar g}=2.4$ line at 
600~nm for instance, the magnetic broadening reaches 2~\kms\ for a 1~kG 
field, similar to the thermal broadening of a Fe spectral line in the 
atmosphere of a main sequence star but smaller than the macroscopic 
turbulence that often widens further the spectral lines of most cool stars. 
A careful analysis of spectral profiles of low-mass stars, comparing in 
particular the shapes of lines with similar formation conditions but 
different magnetic sensitivities, can lead to estimates of the relative 
area of the visible stellar hemisphere covered with magnetic fields (the 
filling factor $f$) and of the average field strength \Bs\ within these 
active regions \citep{Robinson80a, Saar88};  these 2 quantities are 
degenerate to some extent, with the magnetic flux $f \Bs$ being more 
accurately determined than either $f$ or \Bs\ individually.  
In intermediate- and high-mass magnetic stars, fields are not spatially 
intermittent (subsurface convection being either weak or inexistent) and 
more or less cover the whole photosphere (i.e., $f\simeq1$), allowing \Bs\ 
to be estimated directly.  

An extension of this technique to molecular lines (the Wing-Ford band of 
FeH at 0.99~$\mu$m) was recently proposed and applied to a small set of 
very-low-mass dwarfs \citep{Valenti01, Reiners06b}.  The Zeeman effect 
of molecular lines, and in particular their Land\'e factor, is still 
rather poorly documented, with almost no measurements from laboratory 
experiments and only very recent estimates from semi-empirical modelling 
\citep{Afram08}.  As a result, the corresponding technique for measuring 
fields essentially consists at expressing the FeH spectrum of a star 
as a linear combination of those of 2 carefully selected reference stars 
known for their respectively strong (i.e., multi kG) and weak (i.e., undetectable) 
magnetic fluxes;  although rather crude, this method nevertheless provides 
an interesting first-order option for extracting the magnetic information 
coded into FeH lines of low-mass stars.  Improvements to this pioneering 
work are expected soon as laboratory measurements of molecular lines and 
of their magnetic sensitivities become progressively available.  

An obvious way of improving the sensitivities of these various techniques 
is to use spectral lines at nIR wavelengths;  at 2.2~$\mu$m for 
instance, a $\bar g=2.5$ line provides a magnetic splitting of 7.7~\kms\ for a 1~kG 
field, easily detectable in most main-sequence stars with only moderate 
rotational line broadening.  This is obviously mainly interesting for 
low- to very-low-mass stars whose fluxes and spectral-line densities peak 
at red and nIR wavelengths (depending on spectral type) and whose rotation is 
usually slow in average;  it has been successfully applied over the last two 
decades \citep[e.g.,][]{Saar85, Valenti95}.  
The higher sensitivity allows in particular to derive a rough description 
of how field strengths spread at the surface of the star, rather than just 
an estimate of the average magnetic intensity \citep[e.g.,][]{Johns99b};  
in some cases, it is even possible to detect Zeeman splitting directly 
\citep[e.g.,][]{Saar85}.  The nIR also provides a good opportunity for studying magnetic 
fields in cool dark spots of low-mass stars, mostly outshined by the 
surrounding photosphere at visible wavelengths but accessible in the nIR 
where the spot/photosphere contrast is much smaller.  
The advent of new efficient nIR high-resolution \'echelle spectrographs 
in forthcoming years (e.g., GIANO on the Telescopio Nazionale Galileo 
or TNG, and SPIRou on the Canada-France-Hawaii Telescope or CFHT) should 
further boost such applications.  

Although regularly used since the very beginning of solar and stellar 
magnetometry, these techniques underwent a large burst of applications 
when first successfully used for detecting magnetic fields in low-mass 
stars other than the Sun \citep[e.g.,][]{Robinson80}, and in particular 
on young low-mass protostars \citep[e.g.,][]{Johns99b, Johns07}.  
Being insensitive to the field topology, they provide a 
very easy and efficient way of diagnosing tangled magnetic fields for 
which polarisation signals are very weak (see below).  
The drawback is however that the information they yield is rather 
limited;  in particular, they are almost useless at deriving information 
on how the field is oriented and how it splits into its poloidal and 
toroidal components.  

\subsection{Photo- and spectropolarimetry}

The polarisation properties of a Zeeman-split line furnish a second major 
means of measuring magnetic fields.  In particular, polarisation gives 
access to the orientation of the field:  circular polarisation (from 
$\sigma$ components) is sensitive (at first order) to the line-of-sight (or 
longitudinal) component of the magnetic field, while linear polarisation (from 
$\pi$ and $\sigma$ components) gives access to the perpendicular (or 
transverse) component of the magnetic field.  Analysing the relative degree of 
circular and linear polarisation across line profiles requires a polarimeter, 
i.e., an instrument measuring the differential intensity between two orthogonal 
states of polarisation (e.g., right- and left-handed circular polarisation).  

Polarisation in astronomy is usually described using the Stokes vector 
[$I$, $Q$, $U$, $V$].  In this vector, $I$ is the total specific intensity 
of light in the beam.  $Q$ and $U$ describe the linear polarisation of the 
beam, given by $Q=\langle I_{0\deg}-I_{90\deg}\rangle $ and $\langle U=I_{45\deg}-I_{135\deg}\rangle $ where 
$\langle \rangle $ denotes the temporal average and $I_{0\deg}$, $I_{45\deg}$, $I_{90\deg}$ and 
$I_{135\deg}$ correspond to the intensity the beam would have if filtered 
by a perfect linear polariser with its transmission axis respectively set 
to 0\deg, 45\deg, 90\deg\ and 135\deg\ with respect to a reference direction 
(usually the north).  
Similarly, $V$ describes the circular polarisation of the beam, given by 
$V=\langle I_\circlearrowright-I_\circlearrowleft\rangle $, where $I_\circlearrowright$ 
and $I_\circlearrowleft$ correspond to the beam intensities when filtering 
respectively by perfect circular right and circular left polarisers.  
More information about Stokes parameters can be found in \citet{Landi04}. 

At optical and nIR wavelengths, a polarimetric analysis is usually obtained by means 
of a beam splitter (e.g., a linearly birefringent crystal like a simple calcite block, 
or a combination of several like a Wollaston prism), associated with retardation 
devices (e.g., crystalline plates, liquid crystals or Fresnel rhombs) to 
select the appropriate polarisation state.  By yielding two beams with 
respective intensities equal to $I_{0\deg}$ and $I_{90\deg}$, a beam 
splitter (aligned on the polarisation reference direction) provides the most natural 
method of measuring the Stokes $Q$ linear polarisation of an incoming beam.  
By introducing a retarder in front of the beam splitter (e.g., a quarter-wave 
plate oriented at 45\degr\ with respect to reference axis), one can measure the 
polarisation of interest (e.g., Stokes $V$) by converting it into the beam splitter 
reference linear polarisation (Stokes $Q$);  rotating the retarder further allows to 
exchange both beams emerging from the beam splitter, providing a convenient check that 
the beam difference content is indeed polarisation.  

Different flavours of polarimeters have been proposed over the first century 
of solar and stellar magnetometry.  For the first half-century, Hale and 
followers (including Babcock) turned their high-resolution spectrographs 
into spectropolarimeters by simply adding a quarter-wave retarder and a 
beam splitter in front of the spectrograph slit \citep{Hale08, Babcock47}.  
From the wavelength shifts of magnetically sensitive lines recorded in the 
two orthogonal circular polarisation states, one could derive the brightness 
average of the longitudinal field over the observed solar region or the 
visible stellar hemisphere (called \Bl\ hereafter), with relatively poor 
accuracies (of typically 300~G) by modern standards.  

In the 1970's, a new device was proposed by \citet{Angel70}, consisting at 
measuring the degree of circular polarisation in the wings of an H line of 
the Balmer series, usually H$\beta$.  With its dual-beam polarimetric module 
and its twin photometers (one for each polarimetric channel) coupled to 
narrow-band interference filters (to isolate the line wings), this simple design 
shortcuts the need for a spectrograph and provides a very compact and 
easy-to-operate instrument.  With this device, fields of magnetic chemically 
peculiar A and B stars were observed regularly for about two decades, 
providing a wealth of reference longitudinal fields with error bars as low 
as 50~G for the brightest stars \citep[e.g.,][]{Borra80, Landstreet82}.  
A similar technique can be used to observe the transverse component of the 
magnetic field through the linear polarisation signatures it generates in line 
profiles.  Although very small, the signal does not cancel out once averaged 
over wavelengths thanks to differential saturation between the $\pi$ and $\sigma$ 
components \citep{Leroy62}; this makes broad-band linear polarimetry a viable 
tool for monitoring transverse magnetic fields at the surfaces of stars and an 
interesting option for complementing longitudinal field data \citep[e.g.,][]{Leroy95}.  

All first-generation techniques mentioned above essentially derive one single 
magnetic measurement (of either the surface-averaged longitudinal or transverse 
field) per observation.  However, the information available throughout the full 
circular (Stokes $V$) and linear (Stokes $Q$ and $U$) polarisation profiles across 
spectral lines is far richer.  Following \citet{Donati97a}, the surface-averaged 
longitudinal field (i.e., \Bl, in G) can be derived from the Stokes $V$ profile 
with a good accuracy using:  
\begin{equation}
\label{eq:bell}
\Bl = -714 \frac{\int v V(v) dc }{\lambda {\bar g} \int [1-I(v)] dv}
\end{equation}
where $v$ is the velocity shift (in \kms) with respect to the line central 
wavelength $\lambda$ (in $\mu$m).  The longitudinal field being essentially the 
first moment of the Stokes $V$ profile, it retains only the global circular 
polarisation content and smooths out all small-scale polarisation structures.  
While \Bl\ is adequate to characterise simple large-scale fields, it clearly 
misses most of the information for complex field structures, explaining {\em a 
posteriori} why all early attempts to detect the complex magnetic topologies 
of cool stars failed.  

While using higher order moments of the polarisation profiles in addition to \Bl\ 
can certainly help \citep[e.g.,][]{Mathys89}, the best solution for losing no 
polarisation signal is obviously to extract the Stokes profiles themselves rather 
than some integrated (and thus degraded) quantity.  This is achieved by coupling 
a polarimeter to a digital (usually high-resolution) spectrograph, a technique 
called spectropolarimetry.  
Using an \'echelle spectrograph provides the additional advantage of measuring 
Zeeman signatures from hundreds or thousands of spectral lines simultaneously, 
thereby tremendously increasing the overall efficiency of the process; 
however, it requires the polarimeter to be achromatic enough to derive polarisation 
spectra from a minimal number of sub-exposures.  A cross-correlation-type technique 
called Least-Squares Deconvolution \citep[LSD, ][]{Donati97b} was introduced to 
derive average Zeeman signatures from all medium to strong spectral lines available 
in the wavelength domain.  Thanks to LSD, noise levels in the polarimetric signals 
can be strongly reduced, up to a factor of several tens in late K stars, provided 
that the instrument can collect the whole optical domain in a single shot.  

\begin{figure}
\includegraphics[width=7cm,angle=270]{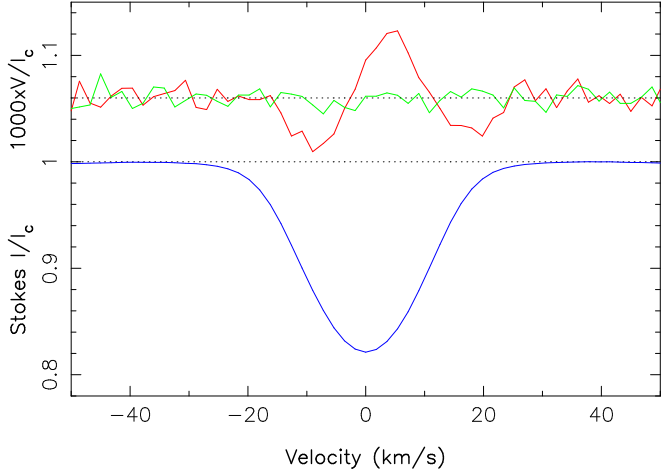}
\caption{LSD circular polarisation (Stokes $V$) Zeeman signature (red line), null polarisation 
check (green line, both expanded by 1000 and shifted vertically by 1.06 for graphical purposes) 
and unpolarised (Stokes $I$) profile (blue line) from the photospheric lines of $\tau$~Boo, 
as derived from ESPaDOnS data.  A clear Zeeman signature is detected (with nothing visible in 
the null polarisation check) with a full amplitude of about 0.01\% of the unpolarized continuum.  
The associated polarimetric sensitivity is 0.001\%, i.e., 10~ppm.   }
\label{fig:lsd}
\end{figure}

In the last two decades, several new spectropolarimeters were developed.  A first 
prototype instrument was tested and validated at the Anglo-Australian Telescope 
\citep{Donati97b, Donati03a}, inspired from the initial concept of \citet{Semel93}.  
A second prototype instrument \citep{Donati99} was installed on the 2m T\'elescope 
Bernard Lyot (TBL) atop Pic du Midi (southwest France), providing the whole community 
with a wider access to spectropolarimetry.  
A new-generation high-resolution spectropolarimeter (called ESPaDOnS, \citealt{Donati03}) 
was installed in 2004 on the 3.6m Canada-France-Hawaii Telescope (CFHT) atop Mauna Kea 
(Hawaii), soon complemented by a clone version (NARVAL) installed at TBL.  
Both instruments are fully optimised for spectropolarimetry, perform a very achromatic 
polarimetric analysis (using modified Fresnel rhombs), and yield full coverage of 
the optical domain (0.37 to 1~$\mu$m) at a spectral resolution of 65,000 and with a 
peak efficiency of about 15\% (telescope and detector included).  
Thanks to their high throughput and fringe-free polarimetric analysis, ESPaDOnS 
and NARVAL can reach very high photon-noise-limited polarisation accuracies (e.g., 
see Fig.~\ref{fig:lsd});  their unprecedented sensitivity (as low as 0.1~G on 
bright narrow-lined cool stars) brings a fresh opportunity to explore magnetic fields 
across most of the HR diagram and already enabled the discovery of magnetic fields in 
several stellar classes not previously known as magnetic (see below).  

With its polarimetric mode, FORS1 on the ESO Very Large Telescope (VLT) atop Cerro 
Paranal (Chile) can serve as a low-resolution (about 2,000) spectropolarimeter.  
While its 150~\kms\ resolution element, it essentially amounts to a Balmer line 
polarimeter for slow rotators (such as chemically peculiar stars), but provides an  
interesting opportunity for investigating large-scale magnetic fields in very rapid 
rotators \citep{Bagnulo02}.  Its giant photon collecting power makes it very efficient 
at exploring magnetic fields in distant stellar clusters and at studying how magnetic 
fields can influence the evolution of early-type stars off the main sequence 
\citep{Bagnulo06}.  

\subsection{Parametric modelling and tomographic imaging of magnetic fields}

Once a magnetic field is detected at the surface of a star, one usually looks 
at whether the detected Zeeman signatures exhibit temporal variability over 
time scales of days and weeks.  In most cases, cyclic variability is 
detected and attributed to a non-axisymmetric magnetic field being carried 
around the star by rotation and viewed under different configurations by the 
Earth-based observer.  By recording time-series of the rotationally modulated 
Zeeman signatures, one can in principle extract information on the parent 
magnetic structure that generates the polarisation signals.  This model 
is usually called the oblique rotator.  

The first attempts at modelling magnetic topologies simply aimed at adjusting 
the rotational modulation of the observed longitudinal field with a simple 
magnetic dipole of polar field strength $B_{\rm p}$, whose axis is tilted 
at an angle $\beta$ with respect to the stellar rotation axis.  This amounts to 
fitting the observed \Bl\ values and corresponding rotational phases $\phi$ with 
the following relation \citep{Preston67}:  
\begin{equation}
\label{eq:pres}
\Bl(\phi) = B_{\rm p} \frac{15+u}{20(3-u)} [\cos \beta \cos i + \sin \beta \sin i \cos 2 \pi (\phi-\phi_0)]
\end{equation}
where $i$ is the angle of the rotation axis to the line of sight, $\phi_0$ the 
phase of longitudinal field maximum and $u$ the linear limb darkening constant.  
This approach usually succeeds at fitting \Bl\ data of moderate precision 
and yields a gross estimate of the large-scale field \citep[e.g.,][]{Borra80} 
but usually fails at matching the detailed modulation whenever very high quality 
data are available and when the field is significantly more complex than a 
dipole \citep[e.g.,][]{Wade00, Donati06b}.  
More sophisticated models were also proposed, e.g., involving a combination 
of non-aligned dipole, quadrupole and octupole terms \citep{Bagnulo96}, in an attempt to fit 
simultaneously constraints from independent data sets (e.g., \Bl\ values and 
broad-band linear polarisation).  Despite the apparent success, detailed 
comparisons with time-series of Stokes profiles on several stars demonstrated 
that this basic model is far too limited and cannot yield a precise
description of the parent magnetic topologies, even in the case of the fairly 
simple fields of magnetic chemically peculiar stars \citep{Bagnulo01}.  

\begin{figure}
\includegraphics[width=12cm]{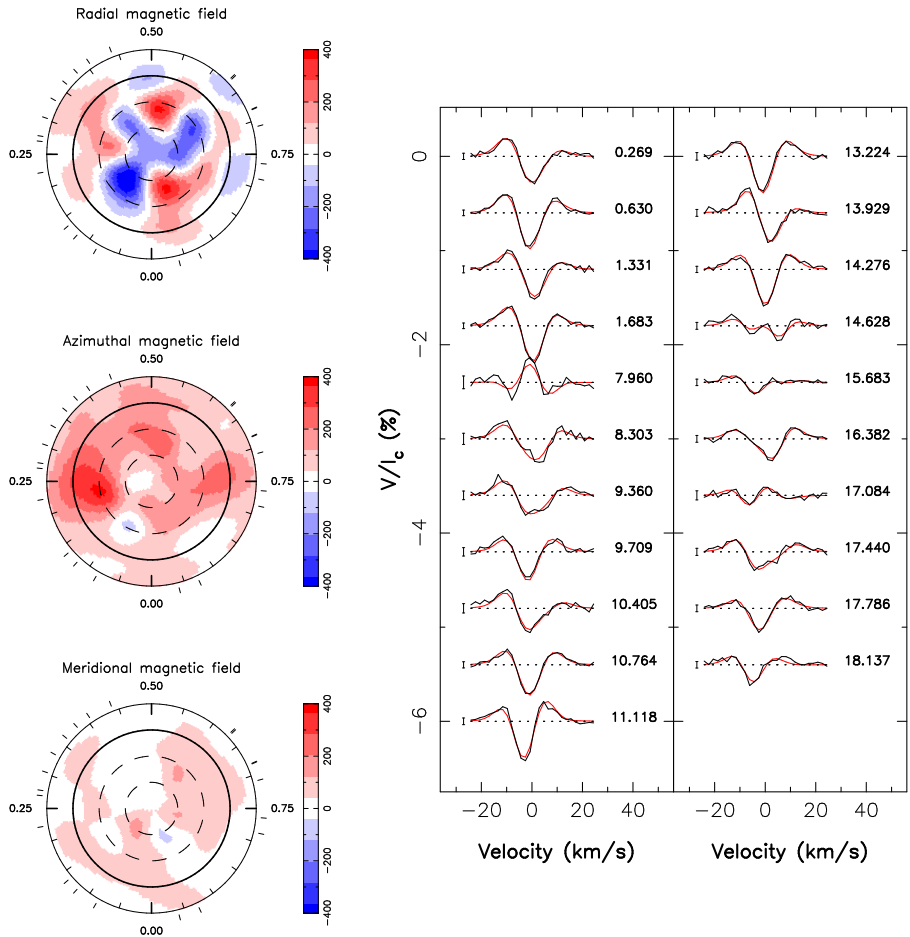}
\caption{Magnetic imaging of the large-scale field of the early-M dwarf DT~Vir using ZDI (left panel) 
from a time series of circular polarisation (Stokes $V$) Zeeman signatures covering the whole 
rotation cycle (right panel).  The reconstructed magnetic topology includes a significant toroidal 
component (showing up as unipolar azimuthal fields over the visible hemisphere) 
and a mostly non-axisymmetric poloidal component, typical of F to early-M dwarfs.  
{\bf Left panel:} the 3 components of the field in spherical coordinates are
displayed (from top to bottom) with magnetic fluxes labelled in G, the star being shown in flattened
polar projection down to a latitude of --30\degr.  Radial ticks around each plot indicate phases
of observations.  {\bf Right panel:}  observed Zeeman signatures are shown in black while the fit
to the data is shown in red.  The rotational cycle and 3$\sigma$ error bars of each observation
are shown next to each profile \citep[from][]{Donati08d}.  }
\label{fig:zdi}
\end{figure}

By modelling the rotationally-modulated Zeeman signatures directly rather than 
a few of their low-order moments only, one can in principle recover to some extent 
(through indirect tomographic techniques like those used for medical imaging) 
the parent magnetic topology.  This technique, combining the advantages of  
Zeeman spectropolarimetry and Doppler imaging \citep[e.g.,][]{Vogt87}, is called 
Zeeman-Doppler Imaging (ZDI) and was first proposed by \citet{Semel89} in the 
particular case of rapidly rotating stars.  The initial implementation of ZDI 
\citep{Brown91, Donati97c}
inverted sets of Zeeman signatures into surface distributions of 
the vector magnetic field, described as 3 series of independent image pixels 
(one series for each component of the magnetic field in spherical coordinates).  
In the latest implementation \citep{Donati01, Donati06b}, the magnetic field is 
decomposed in its poloidal and toroidal components, both expressed as spherical 
harmonics expansions;  this newer method is found to be not only much more robust 
(especially for low-order large-scale fields like dipoles) and more physical 
(for the field description), but also more convenient (e.g., allowing to fine 
tune the respective weight of spatial scales) and informative (poloidal and 
toroidal field components are key ingredients in most theoretical studies on 
magnetic stars, \citealt{Mestel99}).  While most efficient for rapidly rotating 
stars, this method is also applicable to slow rotators, though limited to 
low-order spherical harmonic modes \citep[e.g.,][]{Donati06b, Donati08e}.  
Note that magnetic mapping is practical both for stars with no intrinsic field variations 
and for stars with variable fields, provided that the typical timescale on which 
the field evolves is long compared to the rotation period.  
Numerous results have been obtained with ZDI on all types of stars, from sets of 
LSD Stokes $V$ profiles.  
An example reconstruction in the case of a moderately rotating star is shown 
in Fig.~\ref{fig:zdi};  other examples in the case of rapid and slow rotators 
can be found in the literature \citep[e.g.,][]{Morin08a, Morin08b}.  

By mapping Zeeman signatures over several successive rotation cycles, 
ZDI can also estimate the amount of azimuthal shear (i.e., surface differential 
rotation) that stellar magnetic topologies are subject to.  This method assumes 
a Sun-like surface rotation pattern with the rotation rate varying with latitude 
$\theta$ as $\omeq-\dom\sin^2 \theta$, \omeq\ being the angular rotation rate 
at the equator and \dom\ the difference in angular rotation rate between the 
equator and the pole.  By carrying out magnetic reconstructions (at constant
information content) for a range of \omeq\ and \dom\ values, one can investigate
how the quality of the fit to the data varies with differential rotation.  A 
well defined minimum at physically sensible values of \omeq\ and \dom\ suggests 
that differential rotation is present at the surface of the star 
\citep[e.g.,][]{Donati03b}.  

A variation of ZDI, called Magnetic Doppler Imaging \citep{Piskunov02}, also 
aims at mapping magnetic fields on stellar surfaces from sets of time-resolved 
Zeeman signatures, incorporating polarised radiative transfer in the 
reconstruction process.  This alternate method, applied to sets of $Q$, $U$, and 
$V$ profiles of individual lines proved successful at reconstructing 
detailed magnetic topologies of a few chemically peculiar stars 
\citep[e.g.,][]{Kochukhov04}.  However, using only a few individual lines 
drastically reduces the polarimetric accuracy, limiting in practice the 
applicability of this technique to the brightest and most magnetic stars only;  
moreover, as of today, it does not allow to recover 
directly the poloidal and toroidal components of stellar magnetic fields, 
making it more difficult for comparing observations with theoretical 
predictions.  

\section{Magnetic fields of low-mass stars}
\label{sec:lowm}

\subsection{Activity, rotation and magnetic fields}

Most cool stars exhibit a large number of solar-like activity phenomena; 
dark spots are present at the their surfaces \citep[e.g.,][]{Berdyugina05}, 
where they come and go on timescales ranging from days (as they are carried 
in and out of the observer's view by the star's rotation) to months (as 
they appear and disappear over a typical spot lifetime) and years or 
decades (with spots fluctuating in number and location throughout 
activity cycles).  
Prominences are also detected in cool stars, both as absorption and emission 
transients (e.g., in Balmer lines) tracing magnetically confined clouds 
\citep[e.g.,][]{Cameron89, Donati00} either transiting the stellar disc 
(and scattering photons away from the observer, as for dark filaments on the 
Sun) or seen off-limb (and scattering photons towards the observer, as for 
bright prominences on the Sun).  
Cool stars are also surrounded by low-density coronal plasma at MK temperatures 
showing up at various wavelengths in the spectrum (e.g., radio, X-ray and optical 
line emission) and associated with frequent flaring, recurrent coronal mass 
ejections, and winds.  
Activity phenomena in cool stars scale up with faster rotation and later 
spectral types \citep[e.g.,][]{Hartmann87, Hall08}.  

The current understanding is that activity phenomena are a by-product of
the magnetic fields that cool stars generate within their convective envelopes
through dynamo processes, involving cyclonic turbulence and rotational shearing \citep{Parker55}.
In the particular case of the Sun, dynamo processes are presumably concentrating 
in a thin interface layer (the so-called tachocline) confined at the base of the 
convective zone (CZ) and where rotation gradients are supposedly largest 
\citep[e.g.,][]{Charbonneau05}.
The spectacular images of the Sun collected with TRACE and (more recently) HINODE 
demonstrate that the activity of the Sun very tightly correlates with the presence 
of magnetic field emerging from the surface, either in the form of 
large closed loops (mostly at medium latitudes) or open field lines (mostly at 
high latitudes);  the exact process through which magnetic fields succeed at 
heating the tenuous outer atmosphere to MK temperatures is however still unclear.  

Cool stars are assumed to behave similarly.  This is supported by observations 
showing that activity scales up with rotation rate (at any given spectral type), 
as suggested by dynamo theories.  One of the key parameter for measuring the efficiency 
of magnetic field generation is the Rossby number $Ro$, i.e., the ratio of the rotation 
period of the star to the convective turnover time.  It describes how strongly 
the Coriolis force is capable of affecting the convective eddies, with small 
$Ro$ values indicating very active stars rotating fast enough to ensure that 
the Coriolis force strongly impacts convection.  The observation that activity 
correlates better with $Ro$ than with rotation \citep[e.g.,][]{Noyes84, Mangeney84, 
Kiraga07}, or equivalently, that cooler stars are relatively more active at a 
given rotation rate, agrees well with the theoretical expectation that convective 
turnover times increase with decreasing stellar luminosities.  

Magnetic fields are also responsible for slowing down cool stars through 
the braking torque of winds magnetically coupled to the stellar surface 
\citep{Schatzman62, Mestel99}.  This is qualitatively compatible with the fact 
that most cool stars rotate slowly (like the Sun itself), with the exception of 
close binaries (whose spin angular momentum is constantly refueled from the 
orbital reservoir through tidal coupling) and young stars (which have not had 
time yet to dissipate their initial load of angular momentum).  Magnetised wind 
models yield a good match to the observed distribution of rotation periods in 
young open clusters of ages ranging from several tens to several hundreds of Myr 
\citep[e.g.,][]{Bouvier07b}, further confirming that magnetic fields are likely 
what triggers the spinning down of cool stars as they arrive on the main sequence.  

The main lesson from the solar paradigm is thus that dynamo processes are 
essentially ubiquitous in all cool stars with outer convective layers and generate 
magnetic fields with a high degree of temporal variability at all timescales.  
Extrapolating the solar analogy much further is potentially hazardous;  
in particular, assuming that conventional dynamo models (entirely tailored to 
match observations of the Sun) also apply to cool stars with very different 
convective depths and rotation rates is subject to caution.  In very active 
stars rotating 100 times faster than the Sun for instance, the magnetic 
feedback onto the convection pattern may be strong enough to distort 
theoretical dynamo patterns;  similarly, very-low-mass fully-convective stars 
obviously lack the interface layer where conventional dynamo processes are 
expected to concentrate, but are nevertheless strongly active.  
Magnetic studies of low-mass stars are therefore our best chance for exploring 
the various faces of dynamo processes over a large range of masses and 
rotation rates.  

\subsection{Magnetic properties of cool stars: field strengths, large-scale topologies,  
differential rotation and activity cycles}

The very first estimates of magnetic fields in cool stars other than the Sun 
were obtained by measuring the differential broadening of spectral lines as a 
function of their magnetic sensitivities \citep{Robinson80a, Robinson80}, 
making it possible to derive the first trends on the magnetic properties of low-mass 
stars \citep[e.g.,][]{Saar01}.  These studies find that the 
average surface magnetic strength \Bs\ is, in most cases, roughly equal to the 
equipartition field, i.e., the field whose magnetic pressure balances the 
thermal pressure of the surrounding gas;  only very active stars with rotation 
periods lower than about 5~d (among which fully-convective M dwarfs and 
young low-mass protostars) strongly deviate from this relation 
\citep[e.g.,][]{Johns96, Johns99b, Valenti01b}.  
A similar behaviour is observed in the Sun, where 
fields of moderately active plages are close to equipartition while those of 
active sunspots are stronger by a factor of 2 or more.  This suggests that 
magnetic regions at stellar surfaces progressively evolve from a plage-like 
to a spot-like structure, with flux tubes having increasingly larger 
sizes or being more tightly packed, as activity increases.  

These studies also find that the average magnetic flux $f \Bs$ at the surfaces 
of cool stars increases more or less linearly with $1/Ro$ until it saturates 
at $Ro\simeq0.1$ (corresponding to a rotation period of about 2~d for a Sun-like 
star), with most of the increase being attributable to the filling factor $f$ 
(at least in moderately active stars).  The detection of a saturation regime, 
confirmed with new magnetic flux measurements from molecular lines in M dwarfs 
\citep{Reiners08}, supports the idea that magnetic fields are eventually capable 
of modifying, if not controlling, the convective motions through some feedback 
mechanism;  this may potentially explain in particular why magnetic regions 
at low and high activity levels are morphologically different.

The first detections of Zeeman polarisation signatures from solar-type stars 
\citep[e.g.,][]{Donati97b} and their tomographic modelling with stellar surface 
imaging tools such as ZDI opened up an alternative option for studying dynamo 
processes.  In particular, the medium- and large-scale magnetic fields accessible 
through ZDI, though energetically less important than magnetic fluxes derived 
from Zeeman broadening, are nevertheless optimally suited for checking topological 
predictions of dynamo models on global fields and their potentially cyclic 
variations, to which other methods are insensitive.   

Initial studies, concentrating on a few very active rapidly rotating stars in 
the saturated-dynamo regime brought surprising results.  In particular, they 
demonstrated that strong toroidal fields can show up directly at the stellar surface, 
in the form of monopolar regions of dominantly azimuthal fields or even complete rings 
encircling the star at various latitudes \citep[e.g.,][]{Donati92, Donati97a, Donati03};  
while tori of strong azimuthal fields are likely present in the Sun at the base of the 
CZ (e.g., to account for the non-stochastic arrangement of surface sunspots, known as 
Hale's polarity law), they do not build up at the surface of the Sun - hence the surprise.  
The poloidal components detected on all 3 active stars mainly consist of a significant 
non-axisymmetric term with alternating patterns of opposite radial field polarities.  
Other studies confirmed and amplified these initial results, reporting the 
presence of strong and often dominant toroidal fields at photospheric level 
\citep[e.g.,][]{Dunstone08a}, even in less active stars with longer rotation periods 
\citep[e.g.,][]{Petit05} or earlier spectral types \citep[e.g.,][]{Marsden06}.  
A recent study focussing on main-sequence Sun-like stars with different rotation 
periods suggests that significant surface toroidal fields are detected whenever the 
rotation period is lower than $\simeq$20~d \citep{Petit08}, i.e., $\simeq$25\% shorter 
than the rotation period of the Sun.  

ZDI observations also demonstrated that large-scale 
magnetic topologies of active stars are latitudinally sheared by surface differential 
rotation at a level comparable to that of the Sun \citep{Donati97a}, with the equator 
lapping the pole by one complete rotation cycle about every 100~d (the so-called lap-time, 
equal to 2$\pi$/\dom, where \dom\ is the difference in rotation rate between the equator 
and the pole).  This conclusion agrees with previous results derived from indirect 
tracers of differential rotation \citep[photometric monitoring, e.g.,][]{Hall91}.  
Differential rotation displays a steep increase with earlier spectral 
types, reaching values of 10 times the solar shear or more in late F stars 
\citep{Barnes05, Marsden06, Dunstone08b, Donati08c}.  This trend is independently confirmed from 
observations of spectral line shapes \citep[e.g.,][]{Reiners06a} and suggests that F 
stars with shallow CZs are departing very strongly from solid-body rotation.  
Observations indicate that magnetic topologies remain more or less stable over timescales 
of $\simeq20$\% the lap-time, suggesting that differential rotation is responsible for 
most of the observed mid-term temporal variability. 
Tidal effects in close binary stars apparently have very little impact either on 
magnetic topologies or on differential rotation patterns \citep[e.g.,][]{Dunstone08a, 
Dunstone08b} apart from maintaining a high rotation rate for both system components.  

The major improvement in instrumental sensitivity brought by ESPaDOnS@CFHT and 
NARVAL@TBL made it possible to start surveying the magnetic topologies of cool stars, 
from mid F to late M stars.  It allowed in particular the large-scale field properties 
of M dwarfs to be investigated for the first time on both sides of the full convection 
threshold \citep[presumably occurring at spectral type M4, i.e., at a mass of 
0.35~\msun,][]{Baraffe98}.  Spectropolarimetric monitoring of the rapidly rotating 
M4 dwarf V374~Peg revealed that the star hosts a strong large-scale mostly-poloidal, 
mainly axisymmetric field despite its very short period (0.44~d), high activity level 
and low $Ro$ \citep{Donati06a, Morin08a};  additional observations of active mid-M 
dwarfs further confirmed that dynamo processes in fully-convective stars with masses of 
about 0.3~\msun\ are apparently very successful at generating strong poloidal fields with 
simple axisymmetric configurations \citep{Morin08b}.  

Comparing to partly-convective early M dwarfs reveals that the transition in the 
large-scale field properties is fairly sharp and located at a mass of about 0.4 to 
0.5~\msun\ \citep{Donati08d}, i.e., slightly above the 0.35~\msun\ theoretical 
full-convection threshold.  This sharp transition also coincides with a strong 
decrease in surface differential rotation (with photospheric shears smaller by a factor 
of 10 or more than that of the Sun) and, logically, with a strong increase in the 
lifetime of large-scale fields \citep{Morin08b}.  
Preliminary results on very-low mass stars ($<0.2$~\msun) suggest that the 
situation is even more complex, with some stars hosting very strong and simple 
large-scale fields (like those of mid-M dwarfs) and some others with much weaker 
and complex magnetic topologies (resembling those of early-M dwarfs). 
Observations of a larger sample are needed to clarify the situation but the 
preliminary results already demonstrate that at least some very-low-mass stars 
are capable of producing a strong large-scale axisymmetric poloidal field.  
This conclusion is independently confirmed by the detection of highly-polarised 
rotationally-modulated radio emission from late M and early L dwarfs attributable 
to intense large-scale magnetic fields \citep[e.g.,][]{Berger05, Berger06, 
Hallinan06} through electron cyclotron maser instability \citep{Hallinan08}.  

\begin{figure}
\includegraphics[height=13cm,angle=270]{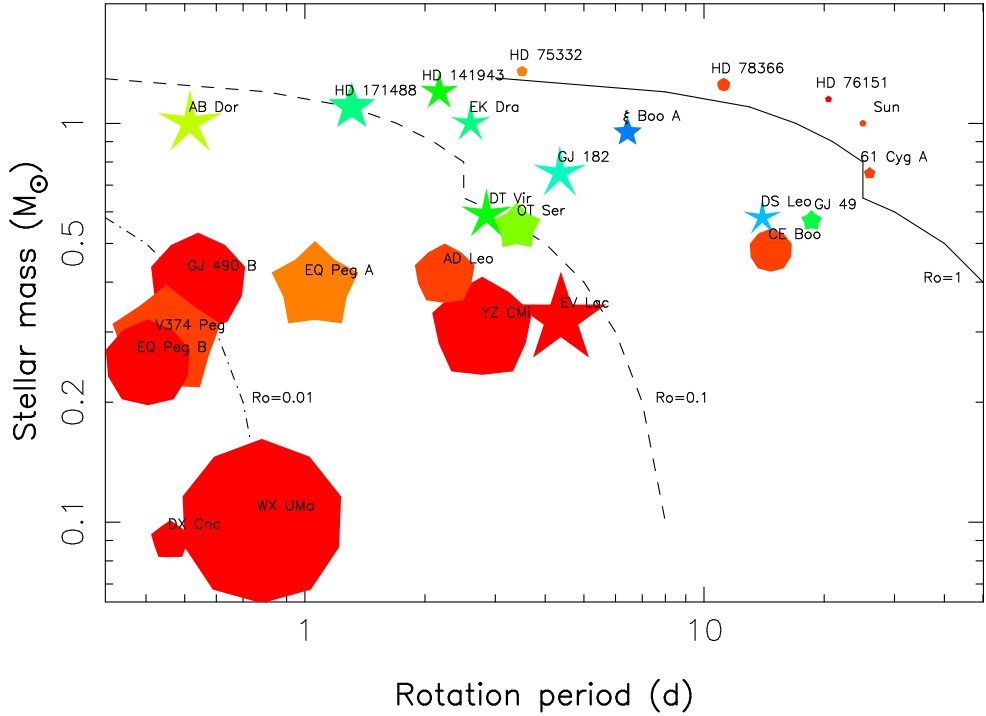}
\caption{Basic properties of the large-scale magnetic topologies of cool stars, as a function
of stellar mass and rotation rate.  Symbol size indicates relative magnetic energy densities $e$,
symbol colour illustrates field configurations (blue and red for purely toroidal and purely poloidal
fields respectively) while symbol shape depicts the degree of axisymmetry of the poloidal field
component (decagon and stars for purely axisymmetric and purely non-axisymmetric poloidal fields
respectively).  The full, dashed and dash-dot lines respectively trace where the Rossby number
$Ro$ equals 1, 0.1 and 0.01 \citep[using convective turnover times from][]{Kiraga07}.
The smallest and largest symbols correspond to mean large-scale field strengths of 3~G and 
1.5~kG respectively.  Results for stars with $\mstar<0.2$~\msun\ are preliminary 
\citep[from][]{Donati08e}. }
\label{fig:mag}
\end{figure}

Figure~\ref{fig:mag} presents graphically the main results obtained so far in the 
framework of the ongoing survey effort, aimed at identifying which stellar 
parameters mostly control the field topology.  To make it more synthetic, the plot 
focusses only on a few basic properties of the reconstructed magnetic topologies, namely 
the reconstructed magnetic energy density $e$ (actually the integral of $B^2$ over 
the stellar surface), the fractional energy density $p$ in the poloidal field 
component, and the fractional energy density $a$ in mostly axisymmetric modes (i.e., 
with $m<\ell/2$, $m$ and $\ell$ being the order and degree of the spherical harmonic 
modes describing the reconstructed field).  Each selected star is shown in the plot 
at a position corresponding to its mass and rotation period, with a symbol depicting 
these three characteristics of the recovered large-scale fields, i.e., $e$ (symbol 
size), $p$ (symbol colour) and $a$ (symbol shape).  This plot clearly illustrates the 
two main transitions mentioned above: 
\begin{itemize}
\item below $Ro\simeq1$, stars more massive than 0.5~\msun\ succeed at producing a
substantial (and sometimes even dominant) toroidal component with a mostly 
non-axisymmetric poloidal component;  
\item below 0.5~\msun, stars (at least very active ones) apparently manage to trigger 
strong large-scale fields that are mostly poloidal and axisymmetric.  
\end{itemize}

\begin{figure}
\includegraphics[width=13.5cm]{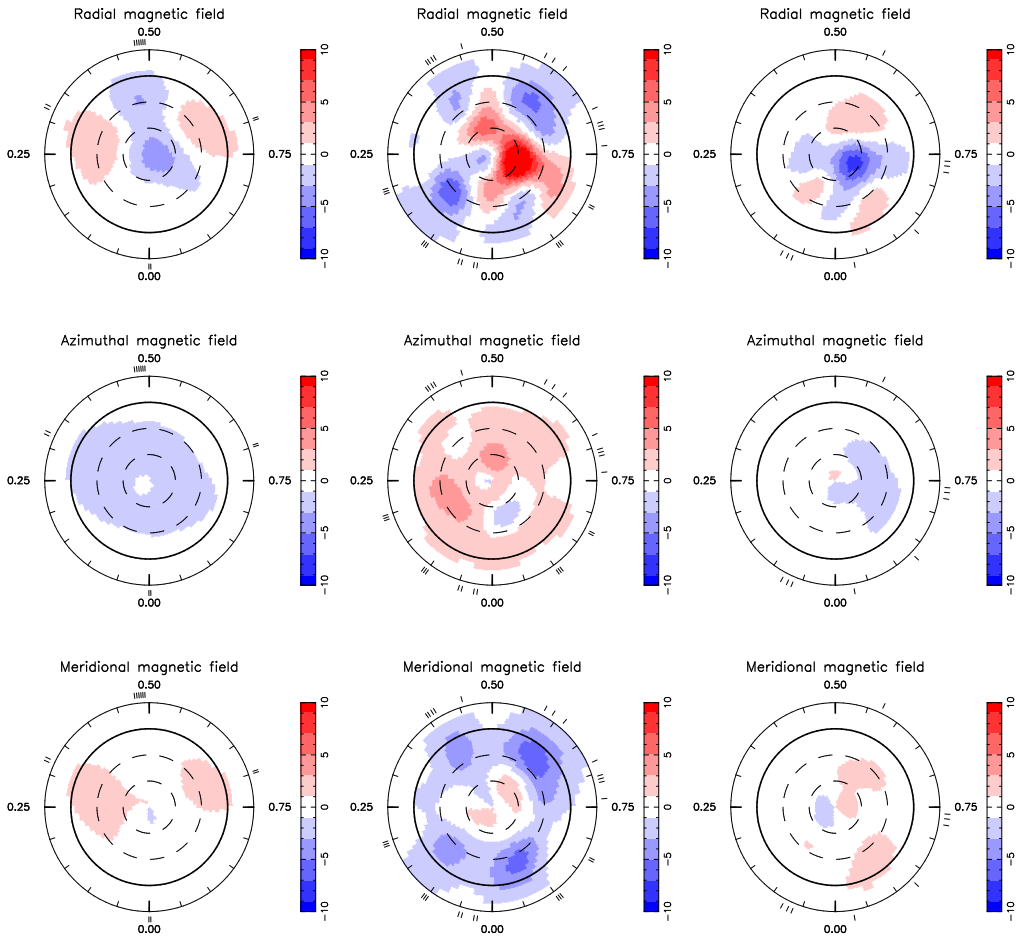}
\caption{Large-scale magnetic topology of the F7 planet-hosting star $\tau$~Boo derived with
Zeeman-Doppler imaging in 2006 June \citep[left panel, from][]{Catala07b}, 2007 June 
\citep[middle panel, from][]{Donati08c} and 2008 June \citep[right panel, from][Far\`es et 
al, in preparation]{Donati08e}.
Plotting conventions are as in Fig.~\ref{fig:zdi}.  Both poloidal and toroidal fields 
globally switched polarities between successive epochs.  }
\label{fig:tauboo}
\end{figure}

Long-term monitoring of large-scale magnetic topologies can potentially reveal whether 
the underlying dynamo processes are cyclic like in the Sun (with the field switching its 
overall polarity every 11~yr), constant or chaotic.  Initial studies carried over a 
decade demonstrated indeed that both the field topologies and the differential rotation 
patterns are variable on long-timescales \citep[e.g.,][]{Donati03a, Donati03b} but have failed 
to catch stars in the process of switching their global magnetic polarities, 
suggesting that their dynamos (if cyclic) do not reverse much more often than that of 
the Sun;  similar conclusions are obtained from long-term monitoring of solar-type 
stars using indirect proxies like overall brightness or chromospheric emission 
\citep[e.g.,][]{Hall08}.
Very recently, first evidence for global polarity switches was reported in a star other 
than the Sun, namely the Jupiter-hosting F8 star $\tau$~Boo \citep{Donati08c}.  
During repeated spectropolarimetric monitoring (see Fig.~\ref{fig:tauboo}), two 
successive polarity switches of 
$\tau$~Boo were recorded within about 2~yr, suggesting an activity cycle about 10 times 
faster than that of the Sun \citep{Donati08e}.  Although still fragmentary, observations 
already show that the poloidal and toroidal field components do not vary in phase across 
the cycle period.  

\subsection{Benchmarking dynamo models with observations of cool stars}

Observational evidence that magnetic fields of cool stars are generated through 
dynamo processes is very strong.  As recalled above, magnetic fields are ubiquitous 
to all stars with significant outer convection (i.e., spectral type later than mid F), 
and direct spectroscopic estimates demonstrate that magnetic fluxes scale up with 
rotation rate (and more tightly with $1/Ro$) until they saturate - in agreement with 
what conventional dynamo theories predict.  
The other option, i.e., that these fields would be fossil remnants from a prior 
evolutionary stage, finds little support either from observations or theory;  while 
models predict fossil fields to be dissipated by convection (as a result of the very 
high turbulent magnetic diffusivity) in as little as 1000~yr \citep[e.g.,][]{Chabrier06}, 
observations 
indicate that fields are very often highly variable (both locally and globally) on 
timescales of months to decades and thus cannot reasonably result from an 
evolutionary process that ended at least tens of Myr before.  
Magnetic field measurements on cool stars thus bring, at first order, strong and 
independent support to generic dynamo models.  

Dynamo models have undergone considerable progress in recent years;  mean-field 
models are now implementing more physics \citep[e.g., the presence of an interface 
layer or the effect of meridional circulation,][]{Parker93, Dikpati99} while 
direct numerical simulations are now able to reach strongly turbulent regimes 
capable of producing intense mean magnetic fields \citep[e.g.,][]{Brun04, Browning08}.  
However, despite such progress, there is still a large number of open questions, 
some of them concerning the very basic physics of dynamo processes, e.g., the 
identification of the primary mechanism through which the large-scale poloidal 
component is regenerated \citep{Charbonneau05}.  Above all, dynamo models are 
almost completely tailored for the Sun, with all model parameters 
finely tuned to reproduce solar observations as well as possible;  checking them 
against observations of other stars with different masses and rotation rates in 
particular is a mandatory validation test that they yet have to undergo.  
The growing body of published results on large-scale magnetic topologies of 
cool stars should provide the opportunity for doing this in the near future.   
Meanwhile, we will summarize here the main topics on which the recent results 
provide new insight into dynamo processes.  

The presence of toroidal fields at the surface of partly-convective stars 
with $Ro<1$ is undoubtedly a surprising discovery, leading some to conclude 
that dynamo processes in very active stars must be operating either throughout 
the whole CZ or at the very least within a subphotospheric layer 
\citep[e.g.,][]{Donati97a, Donati99, Donati03} rather than just at the base of 
the CZ (as usually assumed in conventional dynamo theories).  Interestingly, a 
similar idea - a distributed dynamo shaped by near-surface shear - was 
recently invoked and investigated theoretically in the particular case of the 
Sun as an alternative to conventional interface dynamo models \citep{Brandenburg05};  
in particular, this new model can potentially solve a number of long-standing 
issues (e.g., the large number of toroidal flux belts produced by interface 
dynamos) if further validated by new simulations.  Very recent spectropolarimetric 
observations of the Sun with the HINODE spacecraft revealed that the quiet 
inter-network regions (i.e., the inner regions of supergranular cells of the 
quiet Sun) are pervaded by transient mainly-horizontal magnetic flux \citep{Lites08} 
possibly generated by a near-surface dynamo \citep{Shussler08}, giving still 
further support for a non-conventional distributed and/or near-surface dynamo in 
the Sun.  Admittedly, surface toroidal fields detected in cool active stars are 
more stable than those seen on the Sun and participate to the large-scale field;  
they could however share a similar origin and scale up in strength and size with 
$1/Ro$, being only visible at low ZDI-like spatial resolutions for stars with 
$Ro<1$.  In turn, this may suggest that the newly-discovered horizontal fields 
of the Sun also participate to the activity cycle.  

Despite their high level of activity, fully-convective stars obviously lack 
the interface layer where dynamo processes presumably operate;  understanding their 
magnetism thus represents a major challenge for theoreticians.  
A wide range of predictions has been made about the kind of fields that such dynamos 
can produce;  while early studies speculate that they generate small-scale fields only 
\citep{Durney93}, newer models find that they can potentially trigger purely 
non-axisymmetric large-scale fields \citep{Kuker99, Chabrier06} with CZ's rotating as 
solid bodies \citep{Kuker97}.  The latest simulations show that axisymmetric poloidal fields 
can also be produced with significant differential rotation \citep{Dobler06}, 
but that toroidal fields are usually dominant and differential rotation rather weak
whenever $Ro$ is low enough \citep{Browning08}.  In this context, the recent discovery 
(from both spectropolarimetry and radio observations) that fully-convective stars are 
able to generate strong and simple large-scale mostly-axisymmetric poloidal fields 
while rotating almost as solid bodies \citep{Donati06a, Morin08a, Morin08b} is very 
unexpected and hard to reconcile with any of the existing models.  

The first results of the spectropolarimetric survey indicate that the sharp transition 
in the large-scale magnetic topologies and surface differential rotation of M dwarfs 
more or less occurs where the internal structure of the star drastically changes with 
mass \citep[the inner radiative zone shrinking in radius from 0.5~\rstar\ to virtually 
nothing when \mstar\ decreases from 0.5~\msun\ to 0.4~\msun,][]{Baraffe98, Siess00}.  
It is also worthwhile noting that X-ray luminosities of M dwarfs (relative to their 
bolometric luminosities) are roughly equal (at similar $Ro$) on both sides of the 
full-convection threshold, while the strengths of their large-scale fields feature 
a clear discontinuity \citep[at a mass of about 0.4~\msun,][]{Donati08d}.  
All this suggests that dynamo processes become much more efficient at producing 
large-scale mainly-axisymmetric poloidal fields essentially as a response to the 
rapid growth in convective depths with decreasing stellar masses;  this is 
qualitatively compatible with the idea that the geometry of the CZ may control 
the kind of dynamo wave that a cosmic body can excite \citep{Goudard08}.  

The first detection of global magnetic polarity switches in a star other than the Sun 
is a major first step towards a better understanding of activity cycles of low-mass 
stars.  Looking at, e.g., how cycle periods vary with stellar mass and rotation rate, 
or how poloidal and toroidal fields fluctuate with time across the cycle period, should 
ultimately reveal what physical processes mostly control the cycle.  
Results using a Babcock-Leighton flux transport dynamo model on the Sun 
\citep{Dikpati99} suggest that meridional circulation is a crucial parameter;  while 
meridional circulation is difficult to estimate directly in stars other than the Sun, 
its relation to rotation and differential rotation can potentially be tracked back 
from how cycle periods vary with stellar parameters.  The geometry of the CZ is 
potentially also important \citep{Goudard08}.  

\subsection{Magnetic braking and coronal structure of low-mass stars} 

As mentioned above, the main impact of magnetic fields on low-mass stars is 
to generate extended outer atmospheres including in particular MK coronae and 
magnetised winds; these magnetised winds, as well as the frequent 
ejection of massive coronal prominences trapped in large-scale magnetic loops 
\citep[e.g.,][]{Cameron89, Donati00}, are usually invoked as the main mechanism 
by which low-mass stars efficiently lose their angular momentum and rapidly 
spin down on the early main sequence.  
Most constraints on how fast stars are spinning down as they age come from 
empirical modelling of distributions of, e.g., rotation periods in young open 
clusters of various ages \citep[e.g.,][]{Bouvier07b}, or activity levels and 
line broadening at various ages or vertical distances from the Galactic plane 
\citep[e.g.,][]{Delfosse98, West08, Reiners08b}.  
Results indicate that low-mass stars typically spin down in a timescale of 
about 100~Myr down to about 0.4~\msun;  very-low-mass fully-convective stars 
spin down 10 times more slowly, with brown dwarfs having spin-down times that 
continue to increase as mass decreases.  
This modelling however tells us little about how exactly angular momentum is 
lost and why stars suddenly start to lose much less angular momentum once 
they get fully-convective.  

Modelling coronal structures of cool stars other than the Sun is a promising 
option to address this issue.  This is performed by extrapolating 
large-scale surface magnetic topologies obtained with ZDI up to the 
corona, assuming the field is current-free \citep[e.g.,][]{Jardine02a};  
field lines are also forced to open (e.g., under the coronal pressure) 
at a given distance above the stellar surface from which the field is 
radial (e.g., simulating the base of the wind), with the radial extent, 
temperature and density of the corona being used as free model parameters  
\citep{Jardine02b}.  The free model parameters can be derived by matching 
model predictions to X-ray emission measures and their rotational modulation 
whenever available;  this modelling suggests for instance that X-ray 
emitting coronae of cool stars concentrate fairly close to the stellar 
surface \citep{Hussain07}.  
Applying standard wind models to open field lines derived from 
such three-dimensional magnetic mapping can potentially predict 
angular momentum losses at each spectral type and thus determine 
whether the drastic change in large-scale magnetic topologies reported 
to occur at very-low masses can account for the coincidental 10-fold 
increase of spin-down times.  

\section{Magnetic fields of intermediate- and high-mass stars}
\label{sec:higm}

\subsection{General properties of upper-main sequence magnetic stars}

In contrast to the low-mass stars discussed previously, the intermediate-mass 
(from 1.5 to 8~\msun) and high-mass (above 8~\msun) stars have relatively 
quiescent envelopes, with at most rather shallow CZs (near the surface where 
H, He or He$^+$ is partially ionised).  Instead these stars have strong 
convection in the core where nuclear energy is produced.  As mass 
increases, stellar luminosity grows very rapidly while main sequence lifetime 
shrinks fast (1~Gyr  for 2~\msun, 100~Myr for 5~\msun, and 10~Myr for 
about 17~\msun).  
Unlike their low-mass counterparts, higher-mass stars tend to rotate 
rapidly, i.e., at rotation rates of more than 20\% of the critical 
speed (the rotation rate at which the equatorial velocity equals the Keplerian 
velocity).  Intermediate-mass stars generally appear to have almost no mass 
loss, while massive stars usually lose mass (as a result of their intense 
outflowing radiation field) at such a rate that their subsequent evolution 
is significantly modified.  

Up to now, magnetic fields have only been found among a small minority of 
higher-mass stars.  
For intermediate-mass stars, Babcock's initial discovery \citep{Babcock47} 
of a magnetic field in a star with fairly narrow spectral lines and peculiar 
atmospheric abundances (belonging to the small subgroup of stars called 
``peculiar A/B'' or Ap/Bp stars) has ultimately led to the empirical demonstration 
that all stars of the Ap/Bp class host detectable magnetic fields;  in contrast, 
there are no well-established cases of other intermediate-mass stars with fields.  
These magnetic Ap/Bp stars constitute a few percent of all intermediate-mass 
main-sequence stars of similar spectral types.  
In more massive stars, a number of field detections have recently been 
reported, for example in a few (mostly massive) young Herbig Ae/Be stars and 
early B and O stars, all showing marginal or no signs of chemical 
peculiarities.  These fields are similar in size and structure to those 
found in the Ap/Bp stars and probably represent a higher-mass continuation of 
the same magnetic phenomenon.

In contrast to those of low-mass stars, the magnetic fields of intermediate- 
and high-mass stars often have simple large-scale topologies and exhibit virtually 
no intrinsic variability even on time-scales of decades.  
Furthermore, the observed field strengths do not scale up with rotation rates;  
some of the largest fields occur in stars with rotation periods of months or 
years.  From the positions of Ap/Bp stars with accurate parallaxes and
temperatures in the Hertzsprung-Russell (HR) diagram, and from their location 
essentially on the isochrones of open clusters, it is clear that magnetic Ap/Bp stars
are very similar in bulk structure to normal A and B stars, and their striking 
abundance anomalies seem to be primarily a near-surface effect.   More on general 
properties of magnetic Ap/Bp stars can be found in \citet{Landstreet92}.  

\subsection{Magnetic fields measurements in intermediate- and high-mass stars}

The fields of middle and upper main sequence stars, and their pre-main
sequence precursors, are mostly detected through circular spectropolarimetry, 
which reveals (as in lower-mass stars) the presence of the Zeeman effect in 
spectral lines.  In addition, some of these stars have sufficiently large 
fields and sufficiently small projected rotation velocities that 
Zeeman splitting is visible directly in the intensity (Stokes $I$) spectrum.  
The result of a measurement of circular polarisation is usually described by 
deducing the mean longitudinal field \Bl;  typical values vary over about two
orders of magnitude from one star to another, between about 100~G
and 10~kG;  when Zeeman splitting is visible, the resulting mean surface field 
intensities \Bs\ range from about 2 to 30~kG.  
Additional information on the orientation of transverse magnetic fields in Ap/Bp 
stars is also available through broad-band linear polarisation measurements 
\citep[e.g.,][see also Sec.~\ref{sec:inst}]{Leroy95}.  

The observed fields are usually (though not always) periodically variable, as 
a result of a non-axisymmetric magnetic geometry (about the rotation axis) carried 
around the star by rotation.  Among the magnetic stars that show the specific 
chemical abundance anomalies of Ap/Bp stars, unpolarised spectral lines are often 
also variable with the same rotation period, indicating an inhomogeneous distribution 
of chemical elements over the stellar surface.  This in turn usually leads to small 
periodic photometric variations.  More massive stars also show magnetic and spectrum 
(and sometimes even photometric) variability;  their spectrum variability however 
shows up mainly in lines formed at the base of the wind, indicating that the fields 
of massive stars impact their winds rather than their near-surface distribution of 
chemical elements (as in Ap/Bp stars).  

The observed temporal variations of \Bl\ (as well as of \Bs\ and broad-band 
linear polarisation whenever available) are usually compatible with dipole or 
low-order multipole fields inclined to the rotation axis (the 
oblique rotator model mentioned in Sec.~\ref{sec:inst}), leading early 
studies to conclude that large-scale magnetic fields of Ap/Bp stars are 
globally simple \citep[e.g.,][]{Landstreet82, Leroy95, Bagnulo96}.  
Although rough, this modelling provides a simple way of characterising the 
very-large-scale geometries and strengths of magnetic fields in early-type stars;  
in particular, it enabled to demonstrate that magnetic Ap stars with 
rotation periods of a month or less tend to have magnetic fields perpendicular 
to their rotation axis, while magnetic Ap stars with longer rotation periods 
tend to align their fields on the rotation axis \citep{Landstreet00}.  
More recently, similar modelling applied to a survey of weak-field Ap/Bp stars 
established that their large-scale fields always have a minimum strength of about 
300~G at the surface \citep{Auriere07}; this minimum field, roughly equal to the 
thermal equipartition field in Ap/Bp stars, is apparently necessary to lead 
to the chemical patterns and spectrum variability of the Ap/Bp stars.
Complete time-series of Stokes $V$ profiles (as well as Stokes $Q$ and $U$ 
profiles whenever available) of magnetic Ap/Bp stars collected over full 
rotational cycles \citep[e.g.,][]{Wade00} show that these 
stars also feature a significant amount of medium-scale magnetic structures 
that are not properly described with the simple descriptions first used 
\citep[e.g.,][]{Bagnulo01} but that can be mapped using more sophisticated 
tomographic imaging tools \citep[e.g.,][]{Kochukhov04, Donati06b}.  

Efforts have been made to study empirically the evolution of magnetism
with time through the (rather long) main sequence phase in Ap/Bp stars.
For nearby field stars, for which the magnetic fields are often well
characterised, this is done by using the HR diagram positions with
computed stellar evolution tracks to deduce the ages and fractions of
the main sequence lifetime elapsed for many individual stars, which
are then studied as an ensemble to search for correlations and trends.
The main difficulty with this method is that it requires highly
optimistic (and probably unrealistic) assumptions about the
uncertainties in effective temperature and mass, and about the 
appropriateness of using specific sets of evolutionary tracks for
given bulk chemistries, in order to derive usefully accurate ages for 
stars.  Although various authors have claimed to obtain definite 
conclusions using this method, it is not at all clear that such results 
are meaningful.

Attention has therefore shifted to studying magnetic stars in open clusters, 
a possibility which has only recently opened up with advances in 
spectropolarimetry.  No suitable sample of magnetic cluster members existed, 
and so a large survey was required \citep{Bagnulo06}, which has revealed more 
than 80 magnetic cluster members for which the fields are now roughly 
characterised.  Ages for these stars are determined within about $\pm30\%$, 
making it possible to discern clearly that both fields and magnetic fluxes 
in Ap/Bp stars in the range of 2 to 5~\msun\ decline by a factor of several 
during their main sequence evolution, with much of the decline taking place 
early in the main sequence phase \citep{Landstreet07, Landstreet08}.  

Intermediate- and high-mass magnetic stars have the peculiarity that
they generally rotate more slowly than most normal stars of the same
mass, typically by a factor of five to ten, but sometimes by a factor
of 1000 or more.  Apparently, most of the extra loss of angular momentum 
that magnetic stars experience (with respect to non-magnetic stars) 
occurs during pre-main sequence phases\footnote{Rare cases have been 
reported of magnetic Ap stars changing rotation period by an observable 
amount in recent years \citep[e.g.,][]{Pyper98} through a mechanism 
that is not yet clearly understood.}.  
Another bizarre and mysterious aspect of upper main sequence magnetic
stars is that they are very rarely members of close binary systems \citep{Abt73}, 
although wide binaries are not uncommon.  This fact makes the slow rotation of 
most Ap stars even more striking;  most other slowly rotating A and B stars 
are members of close binary systems. 

Magnetic surveys have been carried out over hundreds of intermediate and 
massive main sequence stars, including a significant number of normal F, A 
and B stars \citep{Landstreet82, Shorlin02, Bagnulo06}, with typical error 
bars on longitudinal fields ranging from 15 to 135~G;  no fields were found 
in any normal stars.  Fields have been sought to no avail in F, A and B stars 
with other chemical peculiarities than those of the Ap/Bp type, such as metallic 
line (Am) stars, mercury-manganese (HgMn) stars, and $\lambda$~Boo stars 
\citep[][with similar error bars]{Bohlender90, Shorlin02, Wade06}.  
The relative frequency of magnetic Ap/Bp stars with respect to all main 
sequence stars of similar mass drops rapidly from a maximum of about 10\% around 
3~\msun\ to zero at about 1.6~\msun\ \citep{Power08}, a fact which has so far 
no explanation.  In solar-type stars, fields are detected up to masses 
of about 1.5~\msun\ (see Sec.~\ref{sec:lowm}) suggesting that there is 
basically no magnetic stars in a narrow mass range above 1.5~\msun.  

Recently, magnetic fields were discovered in several O and early-B 
stars which show little to no chemical abundance peculiarities 
(compared to Ap/Bp stars).  Stars in which fields were found are
relatively slow rotators and usually show periodic variations in spectral 
proxies formed within their strong radiative wind (e.g., X-ray fluxes 
and UV C~{\sc iv} and Si~{\sc iv} lines).  Detecting magnetic fields in 
massive stars is much more difficult than in intermediate-mass ones, in 
spite of the apparent brightness of some.  Such stars have far fewer 
suitable spectral lines in the optical domain available with current 
spectropolarimeters, and most of the available lines are either intrinsically 
broad (lines of H, triplet lines of He), or quite weak (high-excitation lines 
of light elements), or contaminated by emission.  As a result, the threshold 
for significant detection is substantially higher for OB stars than for Ap/Bp 
stars and only 9 reliable detections have been obtained so far, 6 in early B 
stars \citep[namely $\beta$~Cep, $\zeta$~Cas, $\tau$~Sco, $\xi^1$~CMa, Par~1772 
$\nu$~Ori, e.g.,][]{Henrichs00, Neiner03a, Donati06b, VPetit08} and 3 in O 
stars \citep[namely $\theta^1$~Ori~C, HD~191612 and $\zeta$~Ori~A,][]{Donati02, 
Donati06c, Bouret08}.  

\begin{figure}
\includegraphics[width=12cm]{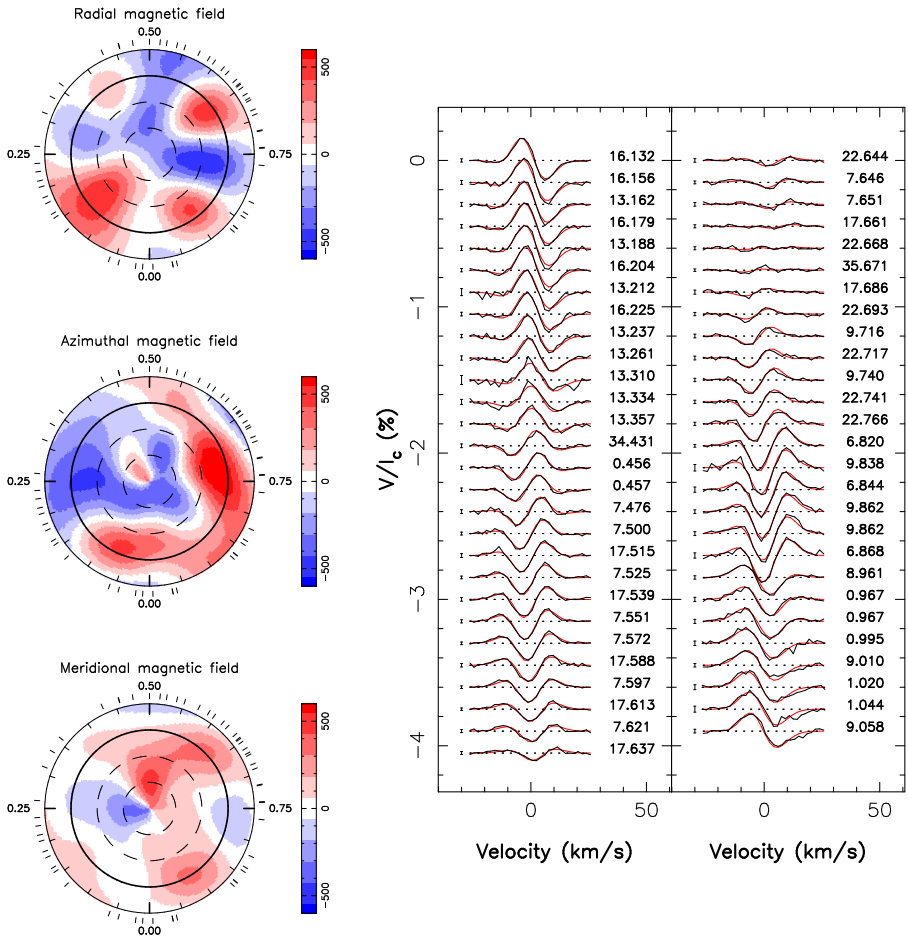}
\caption{Large-scale magnetic topology of the young early B star $\tau$~Sco derived
with ZDI (left panel) from a time series of circular polarisation (Stokes $V$) Zeeman
signatures covering the whole rotation cycle (right panel).  The reconstructed magnetic
field is relatively complex (by the standards of intermediate- and high-mass stars) 
and mostly poloidal.  Both panels are as Fig.~\ref{fig:zdi}.  Note 
the extremely good reproducibility of Zeeman signatures collected at very similar 
phases but very different rotational cycles (rotation period $\simeq41$~d) over a 
total timespan of 4~yr \citep[from][with new material added]{Donati06b}. } 
\label{fig:tausco}
\end{figure}

The magnetic fields of OB stars 
show obvious similarity with the well-known ones of the magnetic Ap/Bp 
stars;  in particular, they are rather simple topologically \citep[except 
in the young B star $\tau$~Sco where the field is significantly more complex 
than usual,][see Fig.~\ref{fig:tausco}]{Donati06b} with global field 
strengths of some hundreds or 
thousands of G;  periodic modulation of spectral or photometric features 
usually correlate well with the magnetic field, and the relative fraction 
of magnetic stars, while still uncertain, is certainly low. 

Another recent discovery concerns magnetic fields of intermediate- and 
high-mass stars as they approach and land on the main-sequence.  
A subset of A and B stars in this situation, known as Herbig Ae/Be 
stars, are usually found in regions of current or very recent star 
formation.  Their positions in the HR diagram place them on 
tracks leading to the main sequence or close to the main sequence itself, 
and their spectra show emission lines usually interpreted as produced 
in (accretion) disks \citep[e.g.,][]{Waters98}.  Initial studies of 
magnetic fields in Herbig Ae/Be stars have yielded a number of 
discoveries \citep[e.g.,][]{Wade07, Catala07, Alecian08a, 
Alecian08b} and already established in particular that only a few percent 
of Herbig Ae/Be stars have significant fields whose properties are 
again very much like those of magnetic Ap/Bp stars.  

Since fields of most magnetic intermediate- and high-mass stars are 
phenomenologically very similar, it is reasonable to suppose that they 
arise from the same origin, with class-to-class differences being due 
to the secondary effects the fields produce on stellar atmospheres, winds, 
and spectra.  A large-scale 4~yr survey (called ``Magnetism in Massive 
Stars'' or MiMeS) using both ESPaDOnS@CFHT and NARVAL@TBL has just begun 
and should ultimately bring a more precise observational description of 
early-type magnetic stars.  

Recently, substantial fields were detected in evolved intermediate-mass stars 
ascending the red giant branch \citep{Auriere08}.  Such stars are cool and 
convective enough to trigger dynamo processes and related activity phenomena 
despite their slow rotation, as suggested by the correlation found between 
their X-ray luminosity and rotation velocities \citep[e.g.,][]{Gondoin99}.  
A small number of these giant stars are expected to be the evolutionary 
descendants of upper main-sequence magnetic stars and thus to show substantial 
fields and enhanced activity;  with a rotation period of over 300~d and a 
field of about 300~G, the late G giant EK~Eri is likely one of them.  

\subsection{Dynamo versus fossil fields}

As discussed in Sec.~\ref{sec:lowm}, the fields of low-mass main sequence 
stars (and of the Sun in particular) are likely generated by contemporary 
dynamo action.  Observations demonstrate that these fields (and all related 
activity processes) are ubiquitous to all cool stars, strongly correlate 
with the presence of a deep-enough outer convective zone, scale up with 
rotation rate and vary on multiple timescales, in 
gross agreement with what theory predicts.  
None of these characteristics are found in upper main sequence stars;  
magnetic fields are present in no more than a small fraction of stars, are 
usually fairly simple topologically, do not strengthen with rotation rate 
(in fact, equally strong fields are often observed in stars having rotation 
rates of days and years), and do not evolve significantly in timescales of 
many years.  In addition, such stars typically have only one or two shallow 
convection zones in their envelopes, vigorous convection occurring only within 
the stellar core.  

Theorists have nevertheless investigated whether the observed fields could 
be due to dynamos operating either in the fully-convective core or in a 
sheared radiative zone, i.e., where the basic ingredients of dynamos 
(turbulence and shear) are present.  Dynamos in fully-convective M dwarfs 
are obviously capable of producing strong axisymmetric fields \citep{Dobler06, 
Browning08} and can presumably do the same in the core of upper main sequence 
stars \citep{Charbonneau01, MacGregor03, Brun05}.  The real difficulty is 
however to bring the magnetic flux to the surface \citep[while not preventing at the 
same time dynamo action to operate;][]{Charbonneau01, MacDonald04} and produce very 
simple and stable magnetic fields with similar properties to those observed in 
early-type stars - in particular the non-correlation of magnetic strengths with 
rotation rates.  
The idea of a more exotic shear dynamo operating within radiative zones of 
early-type stars was also proposed \citep[with the radial shear stretching the field 
into a toroidal configuration that the Tayler/Spruit instability eventually 
destabilises,][]{Spruit02, MacDonald04, Mullan05, Braithwaite06c}.  
Predicted fields are however still expected to 
scale-up with rotation rate and exhibit temporal variability, in contradiction with 
observations.  Last but not least, fields observed in intermediate- and high-mass 
stars are in most cases strong enough to prevent the field from being sheared by 
differential rotation and therefore the instability from operating \citep{Auriere07}.  

In this context, the original theoretical idea, i.e., that the fields of early-type 
stars were generated at some earlier time in the star's history and have been 
retained by magnetic self-induction ever since (without any important current field 
generation taking place), still provides the most convincing picture.  
In this theory, known as the fossil field hypothesis, magnetic fields from the 
interstellar medium thread molecular clouds from which stars form, and are advected 
and amplified as clouds collapse into protostars (see Sec.~\ref{sec:prot} for a
more detailed description);  most of the initial magnetic flux is presumably lost 
(by ohmic dissipation or ambipolar diffusion, see Sec.~\ref{sec:prot}) during the 
contraction process, with only the most magnetic protostars being able to retain a 
significant fraction of their magnetic flux \citep[e.g.,][]{Mestel99} and populating 
the (sparse) class of early-type magnetic stars.  
In these stars, the magnetic field cannot simply decay away through ohmic dissipation.  
Given the high conductivity of their (almost fully ionised) plasma, early-type stars 
have typical ohmic dissipation timescales of order 10~Gyr, i.e., considerably longer 
than their main sequence lifetime.  Any (highly compressed) interstellar magnetic 
field surviving the whole formation process is thus likely to be retained by the 
main-sequence star.  Such fields are expected to be very nearly static, relatively 
simple in structure (and even more so as stars evolve), and not to scale up in strength 
with rotation rate (highly magnetic early-mass protostars actually having more chance 
of being slow rotators, see Sec.~\ref{sec:prot}).  This is in fair agreement with 
what one finds in upper main sequence stars, leading most to conclude that their 
magnetic fields are fossil fields.

The issue of whether simple fossil fields are stable remains however 
open.  Purely poloidal and toroidal fields are known to be unstable 
\citep{Tayler73, Wright73, Braithwaite06b, Braithwaite07};  a mixture 
of both is likely necessary for stabilising a large-scale field.  
Recent numerical experiments \citep{Braithwaite04, Braithwaite06a} suggested 
that many initially unstable global field topologies spontaneously 
develop such a mixed poloidal/toroidal field configuration and become stable 
in doing so.  The apparent existence of a lower limit to the field strengths 
found in magnetic Ap stars \citep{Auriere07} may be related to this very 
instability;  whenever the internal field is too weak, a radial shear 
develops and stretches the field into a predominantly toroidal and unstable 
configuration.  The only magnetic early-type star whose field is 
potentially subject to this instability is the (otherwise normal 
and thus rapidly rotating) O star $\zeta$~Ori~A \citep{Bouret08};  
future observations should reveal whether its magnetism indeed 
results from an exotic shear dynamo capable of generating 
fields in radiative zones.  

\subsection{Diffusion, mass-loss and evolution} 

Magnetic fields can have various effects on early-type stars depending 
on their mass.  

In magnetic Ap/Bp stars, surface abundances are drastically different from 
those of normal non-magnetic stars and obviously correlate with the presence 
of magnetic fields.  With no convection and weak turbulence (e.g., large-scale 
horizontal motions being frozen by the strong large-scale magnetic field),  
microscopic atomic diffusion cause heavy atoms to settle slowly while outward 
radiative acceleration (transmitted to specific ions through their spectral lines) 
forces them to levitate, to rise into the atmosphere and possibly even to 
escape from the star.  Depending on the respective strength of the two forces 
for each species, and on the field orientation, atoms will accumulate 
within (or disappear from) the photosphere, presumably leading to the 
observed persistent chemical anomalies and surface abundance inhomogeneities.  
This scenario fails at reproducing quantitatively existing data 
of Ap/Bp stars, suggesting that it is still lacking some ingredients.  

The more luminous magnetic OB stars show little or no chemical peculiarities, 
a direct consequence of their much higher mass-loss rates.  At about 10~\msun\ 
(i.e., spectral type B2), microscopic diffusion can no longer compete with 
mass loss and no longer influences the surface chemistry significantly 
\citep[e.g.,][]{Michaud86};  the field however is capable of reshaping the
wind by forcing the escaping plasma to follow field lines (at least up to 
the Alfven radius, i.e., wherever the magnetic energy density dominates 
the wind ram pressure).  In this context, wind flows from the two 
magnetic hemispheres collide with each other at the magnetic equator, 
produce a strong shock and generate a corona-like environment trapped at 
the top of closed magnetic loops, with X-ray emitting plasma heated to 
temperatures of up to 100~MK, and massive prominence-like clouds of 
cooling  material corotating with the star \citep{Babel97a, Babel97b}.  

This is indeed observed in early-type magnetic stars, and in particular 
in those massive enough (e.g., $\theta^1$~Ori~C) to power strong winds. 
The model provides a good match to the data whenever detailed observations 
are available \citep{Donati01b, Donati02, Gagne05}.  MHD numerical simulations 
have been carried out for various cases, confirming and extending the predictions 
of the earlier models \citep{UdDoula02, Owocki04, Townsend05, UdDoula06, 
Townsend07, UdDoula08}.  In particular, these computations are able to 
reproduce the anomalously high and hard X-ray emission of magnetic massive stars.  
They also account for the presence and dynamical evolution of corotating 
prominence-like clouds trapped within the magnetosphere of strongly magnetic, 
rapidly rotating Bp stars \citep[initially identified in 
$\sigma$~Ori~E by][]{Landstreet78};  
in this respect, the success of theory at reproducing simultaneously magnetic 
field measurements along with the observed H$\alpha$ and photometric 
variability for $\sigma$~Ori~E itself is outstanding \citep{Townsend07}.  

The slow rotation of most early-type magnetic stars (with respect to the 
non-magnetic ones of similar spectral type) - including those that are just 
approaching the main-sequence - is definite evidence that 
magnetic fields have a drastic impact on the formation of massive stars.  
One possibility is that progenitors of early-type magnetic stars suffer an 
extra loss of angular momentum through a wind or jet, or through interactions 
with their accretion discs \citep{Stepien00};  another promising option is 
that magnetic stars collect less angular momentum from their parent molecular 
cloud during the collapse as a result of their stronger initial magnetic 
flux (see Sec.~\ref{sec:prot}).  This latter mechanism could also potentially 
explain the unusually small fraction of close binaries among Ap/Bp stars.  

Magnetic fields are also expected to impact significantly the evolution 
of massive stars.  Due to their strong angular rotation gradient, 
massive stars are potentially capable of stretching/amplifying their 
magnetic fields, reducing/suppressing their differential rotation and 
modifying their evolution \citep{Maeder03, Maeder04, Maeder05}.  
Massive stars hosting large-scale fields strong enough to resist the 
shear and freeze differential rotation are therefore expected to have 
an even more deviant evolution than those on which models focussed up to 
now.  Evolution of early-type stars off the main-sequence may be studied 
observationally by following stars of a given mass at different 
evolutionary stages \citep[e.g.,][]{Donati06c, Bouret08};  the very 
recent detection of unusually strong large-scale magnetic fields in active 
giants \citep{Auriere08} may similarly hint at how magnetic Ap/Bp stars 
evolve on the long term.   Ultimately, these same magnetic fields may 
reappear as fields of white dwarfs, neutron stars or magnetars. 

\section{Magnetic fields \& star formation}
\label{sec:prot}

\subsection{Quick overview of the star formation process}

In the previous sections, we briefly reviewed the main properties of magnetic 
stars of various masses and the typical phenomena they are subject to, from just 
before they arrive on the main sequence and until they evolve into the giant 
stage;  the strong rotational braking that low-mass stars like the Sun are 
suffering on the early-main sequence is an obvious and famous example.  
These phenomena produce significant effects on evolutionary timescales, both 
in low-mass and massive stars;  however, they remain in most cases a second 
order effect, the magnetic energy of a mature star being always a small 
fraction of its total energy.  The situation is very different in the 
diffuse interstellar medium (ISM) and in the dense cores of giant molecular 
clouds from which stars form, where magnetic, kinetic and gravitational energies 
are roughly comparable to each other \citep[e.g.,][]{Troland86, Crutcher99}.  

As a result of turbulence (inherited from the diffuse ISM), 
giant molecular clouds form clusters of dense self-gravitating condensations 
called prestellar cores (measuring about 0.1~pc across).  
Dense prestellar cores are roughly critical, i.e., their mass is close to the 
magnetic critical mass $\mphi=\Phi/2\pi\sqrt G$ at which the magnetic and 
gravitational energies are equal ($\Phi$ and $G$ being respectively the core 
magnetic flux and the gravitation constant).  
Supercritical dense cores start collapsing while subcritical cores may 
become supercritical through ambipolar diffusion \citep[i.e., with neutral 
gas and dust contracting through the field lines and the ions, 
e.g.,][]{Mouschovias76}.  A rotating accretion disc (hundreds to thousands of 
AU in size) is formed within an accretion shock, usually triggering a powerful 
magnetocentrifugally driven low-velocity molecular outflow \citep[e.g.,][]{Snell80}.  
A protostar progressively forms at the centre of the accretion disc and is 
often associated with a magnetically-collimated jet-like high-velocity 
outflow from the innermost disc regions.   Almost all the magnetic flux and 
angular momentum initially present in the dense core at the beginning of the 
collapse is eventually dissipated, while most of the initial cloud mass 
returns to the diffuse ISM \citep[e.g.,][]{Mestel99}.  

Low-mass star formation is understood best, especially in the later formation 
stages;  at this time (a few Myr after the collapse started), the central protostar 
(called a classical T~Tauri star or cTTS) hosts a large-scale magnetic field strong 
enough to disrupt the inner accretion disc \citep[e.g.,][]{Camenzind90, Konigl91}, 
and to generate a central hole (about 0.2~AU across).  
Accretion from the inner disc rim to the surface of the protostar proceeds through 
discrete magnetic funnels or veils until the disc finally dissipates at an age of 
$\simeq$10~Myr, with the star/disc magnetospheric interaction apparently forcing 
the star into slow rotation \citep{Edwards93, Rebull06}.  Before it dissipates, 
the accretion disc may also form planets.  
While some phenomena are likely common to the formation of low- and high-mass 
stars (e.g., the presence of molecular outflows), some significant differences are 
expected;  for instance, massive stars presumably form in dense clusters of 
highly-turbulent cores, growing quickly in mass and initiating nuclear burning 
while still accreting, with radiation pressure and photoionisation having powerful 
feedback effects on the formation process.  Our present understanding of high-mass 
formation is only fragmentary and poorly constrained by observations 
\citep[e.g.,][]{Zinnecker07}.  

\subsection{Magnetic properties of dense cores, accretion discs and protostars}

Direct magnetic measurements constraining models of star formation 
are difficult and rare.  For the moment, they mostly concern only three stages 
of the formation process: the dense cores of molecular clouds, the protostellar 
accretion discs and the cTTSs.  

As for stars, the Zeeman effect is the only available technique for measuring 
directly magnetic fields in molecular clouds.  From circular polarisation signatures 
in atomic or molecular lines at radio frequencies \citep[typically H~{\sc i}, OH 
and CN lines, e.g.,][]{Troland86, Crutcher99, Crutcher08}, one can derive an estimate 
of the magnetic flux within the cloud with an expression very similar to Eq.~\ref{eq:bell}, 
yielding accuracies on the longitudinal field component of order a few $\mu$G for lines 
like the 1420~MHz (21~cm) H~{\sc i} line or the 1667~MHz (18~cm) OH line.  
Linearly polarised thermal emission from elongated dust grains (with their 
short axis usually aligned along field lines) can also be used to probe the 
magnetic field morphology in molecular clouds and provide an indirect estimate of 
its strength \citep[from estimates of its small-scale randomness of 
orientation,][]{Chandra53, Houde04}.  

Observations are however difficult and results remain sparse, in particular in the 
dark cloud cores that presumably probe best the very early stages of stellar formation.  
Actual Zeeman detections are obtained in only about 2 dozen clouds \citep{Crutcher99, 
Troland08}.  Moreover, only the longitudinal component of the field is detected, 
implying that the derived mass-to-flux ratio (characterising whether or not the cloud 
is magnetically critical) is overestimated;  a statistical correction is possible 
for a large sample of clouds with random magnetic orientations.  The result is that 
dense cores are on average only slightly supercritical with mass-to-flux ratios of 
about 2, confirming that magnetic fields are energetically important in star formation. 
Magnetic maps derived from linearly polarised emission indicate that the magnetic 
field lines are rather regular (with only limited small-scale orientation dispersion), 
suggesting that the cloud cores are roughly critical, in agreement with Zeeman 
measurements;  they also sometimes show a conspicuous hourglass morphology 
\citep[e.g.,][]{Girart06}, indicating that the bending of field lines may provide 
extra support to the cloud and delay its collapse.  
Average magnetic strengths in clouds are observed to scale up with number densities 
$n$ as $n^{0.47}$ above densities of about $10^3$~\pcc\ \citep{Troland86, Crutcher99}, 
providing evidence that the cloud contraction is not spherical and significantly 
influenced by the field.  A very recent study, exploring how the mass-to-flux ratio 
varies from the core to the envelope of the cloud, suggests that the envelope is 
significantly more supercritical than the core \citep{Crutcher08}.  

Magnetic fields are also detected in protostellar accretion discs, i.e., at a later 
stage of the formation process.  Masers from different species (usually OH, H$_2$O, 
methanol) commonly occur in association with high-mass protostellar objects and are 
thought to trace the surrounding circumstellar discs and associated outflows.  
Polarisation measurements indicate the presence of magnetic fields of order a few mG 
at typical distances of 1000~AU from the central object;  magnetic intensities 
again scale with number densities as $n^{0.47}$ \citep[extrapolating the relation 
derived from cloud cores to densities of up to $10^9$~\pcc,][]{Vlemmings08} suggesting 
that the field is still partly coupled to the gas at these more evolved phases of 
the collapse.  In a few properly oriented (i.e., edge-on) objects, the longitudinal 
magnetic field derived from Zeeman signatures switches sign on opposite sides of the 
object, possibly suggesting a disc magnetic field with a mostly toroidal orientation 
\citep[e.g.,][]{Huta05}.  
Paleomagnetic records from meteorites suggest typical magnetic strengths of 0.1--10~G 
at a distance of a few AU for low-mass protostars;  with their randomly organised 
magnetisations, chondrules (whose parent bodies are believed to originate in the 
asteroid belt) are indeed thought to record magnetic fields that predate accretion 
\citep[e.g.,][]{Shu07}.  

\begin{figure}
\includegraphics[width=9cm]{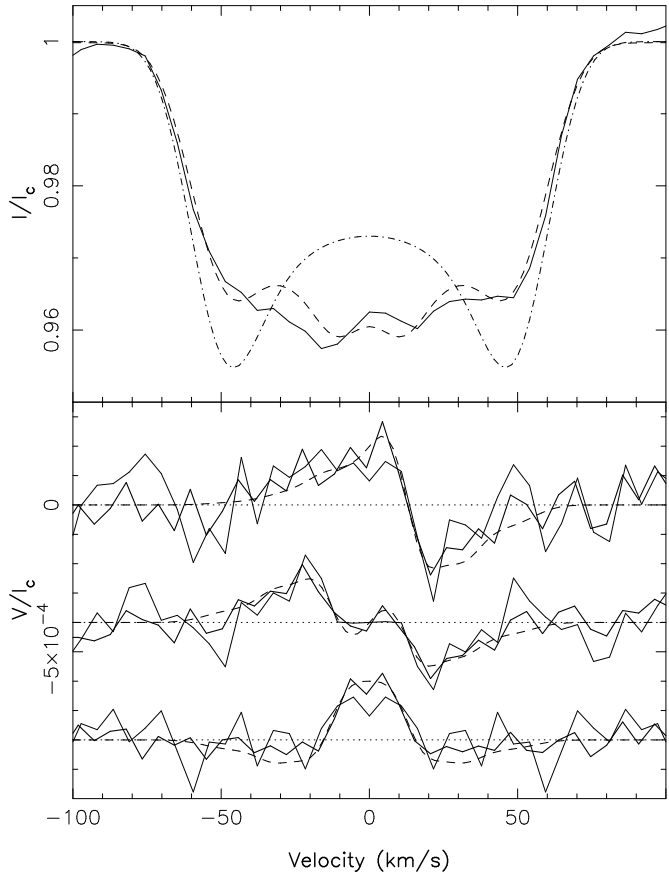}
\caption{Unpolarised and circularly polarised profiles of the protostellar accretion 
disc FU Ori.
{\bf Top panel}: observed Stokes $I$ profile (solid line) and model profiles assuming either a 
Keplerian disc (dash-dot line) or a non-Keplerian disc (with 20\% of the plasma 
rotating at strongly sub-Keplerian velocities, dashed line).
{\bf Bottom panel}: observed Zeeman signature (top curve) split into its antisymmetric and
symmetric components (middle and bottom curves, shifted by $-4$ and $-8\times10^{-4}$)
respectively characterising the vertical and azimuthal axisymmetric magnetic fields.
The model (dashed line) requires the slowly rotating disc plasma to host a 1~kG vertical
field plus a 0.5~kG azimuthal field \citep[from][]{Donati05}.  }
\label{fig:fuori}
\end{figure}

A strong magnetic field was recently detected using optical spectropolarimetry in the 
innermost regions of an accretion disc around a low mass protostar 
\citep[FU~Ori,][]{Donati05} in a supposedly early cTTS stage (with an age of about $10^5$~yr).  
The detected field has a strength of about 1~kG at a distance of only 0.05~AU (where 
the number density is estimated to be $10^{17}$~\pcc) and is found to concentrate in 
the $\simeq$20\% of the disc plasma that rotates at strongly sub-Keplerian velocities.  
From the shape of the Zeeman signature, the orientation of the magnetic field of 
FU~Ori (assumed axisymmetric given the low level of rotational modulation) 
is found to be mainly perpendicular to the disc plane and to include a smaller 
azimuthal component (see Fig.~\ref{fig:fuori}).  

Among all protostellar objects, cTTSs are those on which we have most information 
thanks to their visibility at optical wavelengths.  In particular, magnetic 
fields of cTTSs are well documented, in comparison with those of cloud cores and 
accretion discs.  Magnetic fields with typical strengths of 1--3~kG are measured 
at the surfaces of most cTTSs from Zeeman broadening of (mostly nIR) unpolarised 
line profiles \citep[e.g.,][]{Johns99b, Johns07}.  These magnetic intensities are 
often much larger than thermal equipartition, as in very active cool dwarfs.  Magnetic 
strengths are found to correlate poorly with predictions from current magnetospheric 
accretion models in which fields are assumed to disrupt the inner regions of the 
accretion disc and to ensure approximate corotation between the inner disc rim 
and the stellar surface \citep{Johns07};  this discrepancy is however not 
surprising, Zeeman broadening methods being sensitive to small-scale fields 
whereas predictions of magnetospheric accretion models concern the large-scale 
field.  

\begin{figure}
\includegraphics[width=12cm]{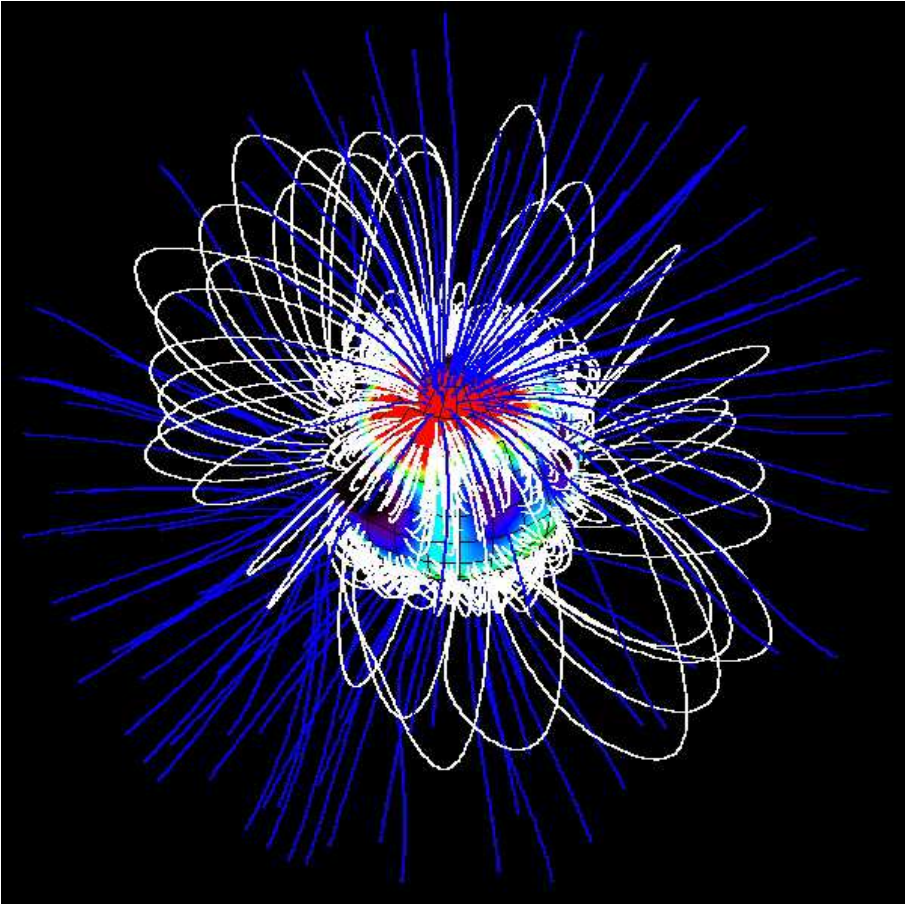}
\caption[]{Magnetosphere of the cTTS V2129~Oph, extrapolated from the surface magnetic 
map derived from spectropolarimetry of both photospheric lines and accretion proxies.  
Open and closed field lines are shown in blue and white respectively, while colours at the 
surface of the star depict the radial field component, with red and blue corresponding to 
positive and negative polarities \citep[from][]{Donati07, Jardine08}. }
\label{fig:v2129}
\end{figure}

Large-scale magnetic fields are also detected with spectropolarimetry.  Circularly 
polarised Zeeman signatures from emission lines (e.g., the He~{\sc i} $D_3$ line 
at 587.6~nm, or the Ca~{\sc ii} infrared triplet at 850~nm) probe magnetic fields 
at the footpoints of the accretion funnels linking the star to the inner disc rim 
\citep[e.g.,][]{Johns99a, Valenti04, Symington05}, whereas Stokes $V$ signatures from 
photospheric lines trace the large-scale magnetic fields permeating the non-accreting 
fraction of the stellar surface \citep{Donati07}.  Average longitudinal fields 
from accreting regions typically reach several kG and display a smooth and simple 
rotational modulation, suggesting a simple large-scale magnetic geometry;  
however, longitudinal fields from the quiet photosphere rarely exceed a few 
hundred G as deduced from complex Zeeman signatures, and apparently trace a more 
tangled parent topology.  Modelling Zeeman signatures from photospheric lines and 
accretion proxies simultaneously (whenever available) reveals that the large-scale 
field is indeed significantly more complex than a dipole (e.g., see 
Fig.~\ref{fig:v2129}) and includes in particular a strong 
octupole component \citep{Donati07, Donati08b}.  In the 2 cTTSs for which a 
magnetic map has been published (namely the 1.4~\msun\ and 0.7~\msun\ 
cTTSs V2129~Oph and BP~Tau), the dipole component of the large-scale field is much 
weaker when the star is not fully convective, as in main-sequence M dwarfs 
\citep{Morin08b, Donati08d}.  

Extrapolating the large-scale magnetic maps of cTTSs to the whole magnetosphere 
allows to estimate the typical distance at which a protostar magnetically connects to 
its accretion disc, and to provide independent constraints on where the inner 
accretion disc rim is located.  Producing high-latitude accretions spots in 
a magnetic topology with a dominant octupolar field at the stellar surface
requires the disc plasma to be injected into the magnetosphere from a distant 
region where the dipole field dominates, i.e., at distances of at least 5~\rstar\  
(or 0.06~AU) in the particular case of the partly convective cTTS V2129~Oph 
\citep{Donati07, Jardine08, Gregory08, Mohanty08}.  This result agrees with 
independent estimates of magnetospheric gap sizes from nIR interferometry 
\citep[e.g.,][]{Pinte08}.  

Information on how magnetic topologies depend on the fundamental parameters of the 
accreting protostar (in particular mass, rotation rate and age) as well as on the 
accretion and outflow properties is still lacking and requires a detailed study on 
a large enough sample of cTTSs;  this is the aim of the Large Program called 
Magnetic Protostars and Planets (MaPP) recently initiated using ESPaDOnS@CFHT and 
NARVAL@TBL.  Little information yet exists on magnetospheric 
accretion processes of more massive protostars (for instance the Herbig Ae/Be 
stars);  while indirect proxies suggest that similar processes may also be 
at work and that magnetic fields could also be involved in at least a small 
fraction of massive stars \citep[e.g., the UX~Ori-type objects,][]{Muzerolle04}, 
conclusive and direct evidence such as that available for cTTSs is still lacking.  

\subsection{Magnetised collapse of molecular clouds}

Until recently, star formation was more or less understood in the framework of the 
standard magnetic model, where magnetic fields control the formation and evolution of 
molecular clouds - including the formation of cores and their gravitational collapse to 
form protostars \citep[e.g.,][]{Mouschovias87, Shu87, Basu97, Li97}.  
In this model, neutral gas and dust contract gravitationally through 
the field and the ions (by ambipolar diffusion), triggering the collapse (when the 
core becomes supercritical) and leaving most of the magnetic flux behind in the envelope.  
However, in recent years a new picture has been proposed, suggesting that turbulence 
(rather than magnetic fields) controls the formation of clouds and cores 
\citep[e.g.,][]{Padoan02} - with clouds forming at the intersection of turbulent 
supersonic flows in the ISM and locally collapsing into cores and protostars wherever 
dense enough to be self-gravitating;  in this model, magnetic fields (while potentially 
present) are too weak to be energetically important.  

The observed scaling law for magnetic strengths with respect to densities 
\citep[$B\propto n^{0.47}$, e.g.,][]{Crutcher99} better agrees with analytical 
ambipolar diffusion models \citep[e.g.,][]{Li97, Basu97};  however, the mass 
distribution of prestellar cores that 
the standard model predicts does not match observations \citep[e.g.,][]{Andre08}.  
While the magnetic paradigm implies longer cloud core lifetimes than the turbulent 
paradigm, they are both roughly compatible with observations \citep[indicating 
intermediate core lifetimes of a few free-fall times, e.g.,][]{Andre08}.  
The very tangled field topologies that the turbulence model predicts in cloud 
cores mostly contradicts observations, field lines being rather regular with 
limited small-scale orientation dispersion;  the very recent result suggesting 
that dark cores are less supercritical than their envelopes \citep{Crutcher08} 
would however challenge the standard magnetic picture if confirmed.  
Given these constraints, theoreticians are now exploring new models in which 
both magnetic fields and turbulence play a significant role in the formation 
of cloud cores \citep[e.g.,][]{Andre08}.

The detection of molecular outflows and collimated jets from protostellar objects 
\citep[e.g.,][]{Snell80} indirectly demonstrates that magnetic fields actively 
participate in the subsequent phase of the cloud collapse.  Magnetocentrifugal models 
were proposed soon after as a tentative explanation;  in these models, the 
collapsing cloud pinches and twists the primeval field into an helical magnetic 
structure very efficient at firing outflows and jets \citep[thanks to the 
magnetocentrifugal force and/or the toroidal magnetic pressure, e.g.,][]{Pudritz83, 
Ferreira97}.  Detailed simulations have been carried out in recent years 
primarily for low-mass stars \citep[e.g.,][]{Machida04, Banerjee06, Hennebelle08a, 
Mellon08}, and in a few cases for massive stars as well \citep[e.g.,][]{Banerjee07}, 
mostly using numerical techniques such as adaptive mesh refinement or nested grids 
to model properly the wide range of spatial scales and densities involved in the 
computations.  The formation steps include the isothermal collapse of the cloud 
core into a flattened structure (the accretion disc) up to the formation of an 
adiabatic core (the first core), the adiabatic collapse within the first 
core, a second isothermal collapse occurring within the first core (triggered by 
the dissociation of molecular hydrogen) and the formation 
of a second core (the protostar itself).  Collapse simulations are able to reproduce both 
the large-scale low-velocity outflows (from the outer cloud regions) and the 
highly-collimated high-velocity jet (from the innermost cloud regions) that 
observations conspicuously show \citep{Banerjee06, Pudritz07, Hennebelle08a, 
Machida08a}.  

Observations unambiguously show that most of the angular momentum (and magnetic 
flux) initially present in the cloud core is dissipated in the collapse, presumably 
by magnetic effects \citep[e.g.,][]{Mestel99, Machida07}.  Simulations indicate that, 
for slightly supercritical clouds,  the magnetic collapse occurs primarily along the 
field lines, making the collapsing envelope denser and flatter than in the non-magnetic 
case;  moreover, no centrifugally supported disc is apparently able to form 
\citep[less angular momentum being delivered to the inner parts and significant angular 
momentum being expelled through magnetic braking,][]{Hennebelle08a, Mellon08}.  
The detection of magnetised plasma rotating at strongly sub-Keplerian velocities in 
the innermost regions of FU~Ori \citep{Donati05} may be evidence that this is 
indeed what happens.  Simulations further suggest that clouds with higher magnetic 
fields may form first cores (and presumably protostars as well) containing less angular 
momentum \citep[e.g.,][]{Hennebelle08a} and are less prone to fragmentation 
\citep[e.g.,][]{Machida08b, Hennebelle08b}.  

The magnetic field detected in FU~Ori corresponds to magnetic fluxes of order a few 
hundred G \citep[1~kG threading about 20\% of the disc,][]{Donati05} at typical number 
densities of $10^{17}$~\pcc, in surprisingly good agreement with the $B\propto n^{0.47}$ 
power law derived from magnetic clouds.  If confirmed, this result suggests that magnetic 
fluxes can potentially survive at densities much higher than initially predicted;  
in particular, this would indicate that yet unidentified ionisation mechanisms 
may operate within protostellar accretion discs \citep[e.g.,][]{Shu07}.  
The field orientation in FU~Ori (compatible with predictions of MHD collapse 
simulations, \citealt{Ferreira97, Banerjee06}, but not with disc-dynamo 
models, \citealt{Brandenburg95, vonRekowski03}) further suggests that 
fields of protostellar accretion discs are of primordial origin.  

The role of magnetic fields in the formation of massive stars is still poorly 
documented, with little direct observations and few numerical simulations of the 
magnetic collapse yet available \citep[e.g.,][]{Vlemmings08, Banerjee07}.  
Observations of intermediate- and high-mass magnetic stars on and immediately 
before the main sequence indicate that these stars rotate much more slowly and 
form close binaries much more rarely than their non-magnetic equivalents, a 
likely result of a different formation process (see Sec.~\ref{sec:higm}).  
This is qualitatively similar to what magnetic collapse simulations predict for 
low-mass stars, with clouds having larger initial magnetic fluxes forming in 
average first/second cores rotating slower and being more often single 
\citep[e.g.,][]{Machida08b, Hennebelle08a, Hennebelle08b};  it is however 
unclear yet how much of this can be extrapolated to more massive stars.  

\subsection{Magnetospheric accretion, angular momentum regulation and protoplanet formation}

Once formed at the end of the second collapse, low-mass protostars host 
strong large-scale magnetic fields whose origin, though not fully clear yet, 
is likely attributable to dynamo processes.  
Any fossil field that survived the collapse is indeed unlikely to survive 
\citep[for more than typically 1,000~yr, e.g.,][]{Chabrier06} the fully 
convective phase that low-mass stars undergo.  The topological similarity of 
cTTSs large-scale fields with those of main-sequence M dwarfs further strengthens 
this conclusion.  Moreover, the lack of large-scale magnetic fields in most 
intermediate-mass protostars \citep[e.g., in Herbig Ae/Be stars,][see also 
Sec.~\ref{sec:higm}]{Wade07} is additional evidence that fossil fields from the 
ISM \citep[while potentially still present in the inner regions of protostellar 
accretion discs,][]{Donati05} eventually perish within most protostars.  

These large-scale magnetic fields strongly impact the life of cTTSs by forcing 
them to interact with their accretion discs;  they evacuate the central disc 
regions, connect the protostars to the inner disc rim, confine the accreting 
material into discrete funnels or veils and slow down the rotation rate of the 
protostar \citep[e.g.,][]{Camenzind90, Konigl91, Cameron93, Shu94, Bouvier07}.  
Several theoretical models and numerical simulations have been proposed and carried 
out to study this complex magnetospheric interaction and attempt to reproduce 
observations, in particular the typical sizes of magnetospheric gaps and rotation 
periods of cTTSs.  Initially restricted to dipolar magnetospheres \citep{Romanova02, 
vonRekowski04, Bessolaz08}, simulations and models now incorporate more complex 
fields resembling those derived from observations \citep[e.g.,][]{Gregory06, 
Long08, Mohanty08, Gregory08}.  Such studies are able to reproduce the presence 
of accretion funnels linking the disc to the star and of accretion spots at 
funnel footpoints, whose locations and geometries are found to depend strongly 
on the inclination of the magnetosphere with respect to the rotation axis of both 
the disc and the protostar \citep[e.g.,][]{Romanova03, Romanova04} as well as on 
the large-scale topology of the field \citep[e.g.,][]{Gregory06, Long08}.  
The size of the magnetospheric gap is however still difficult to reconcile with 
observed large-scale field strengths \citep{Bouvier07, Bessolaz08, Gregory08} and 
simulations fail to confirm that the magnetic torque from the accretion disc is 
strong enough to spin the star down as observations suggest 
\citep[e.g.,][]{Bessolaz08}; a braking contribution from an accretion-powered 
magnetised stellar wind may help solving the problem \citep{Matt05}.  

Magnetic fields in accretion discs are also expected to impact the formation and 
migration of protoplanets.  
In particular, the discs fields and associated MHD turbulence can potentially inhibit 
disc fragmentation through gravitational instabilities and the subsequent formation 
of giant planets \citep{Fromang05};  they can modify as well the migration rate and 
angular momentum of protoplanets \citep{Fromang05b, Machida06}.  
By disrupting the inner regions of accretion discs, magnetic fields of cTTSs 
may stop the inward migration of giant planets formed earlier in the outer disc 
\citep[which would no longer experience the gravitational torque from the 
disc once they enter the magnetospheric gap,][]{Romanova06};  orbital distances of 
most close-in giant planets discovered in the last decade around main sequence stars 
(smaller than 0.1~AU) are indeed compatible with typical sizes of magnetospheric gaps 
in cTTSs.  

\section{Conclusion}

Starting from how magnetic fields of non-degenerate stars can be detected and 
measured and how their large-scale topologies can be reconstructed, we have 
reviewed most observational results available to date and described their 
implications for our understanding of where magnetic fields originate
and how they impact the formation and evolution of stars at different ages 
and for different masses.  

While the first detection of a magnetic field in a star (the Sun) was obtained 
over a century ago, progress in the field has been rather slow until the last 
few decades when the advent of new techniques and instruments made it possible to 
unveil the magnetic fields (and particularly the magnetic topologies) in a wide 
sample of stars throughout the HR diagram.  Observations have revealed 
that magnetic fields strongly influence the formation of stars 
and are ubiquitous to low-mass stars like the Sun;  while 
only present in a small fraction of intermediate- and high-mass stars, magnetic 
fields are nevertheless found to have a profound signature on their atmospheres 
and winds and are able to modify their long-term evolution significantly.  
This progress has triggered a wealth of theoretical studies involving magnetic 
fields, and predictions from many detailed MHD simulations
can now be checked directly with observations and be used to devise 
new modelling ideas and observational tests.  

New instruments, either in construction or in design phase, should further boost 
this field of research in the coming decade.  The Atacama Large Millimeter Array 
(ALMA, operational in 2011--2013) will soon provide direct magnetic measurements 
in various types of protostars at early formation stages and down to spatial scales 
of about 10~AU (thanks to its high angular resolution).  The nIR spectropolarimeter 
SPIRou (a nIR, high radial-velocity-accuracy version of ESPaDOnS and NARVAL 
proposed as a new CFHT instrument for implementation in 2015) should give the 
opportunity of exploring for the first time the magnetic properties of obscured 
protostellar objects and very cool dwarfs;  for protostellar accretion discs 
in particular, SPIRou should be able to access the innermost regions, thus 
nicely complementing magnetic measurements in the outer regions obtained 
with ALMA.  

These new observations, along with updated models and simulations, should ultimately 
bring a much clearer view of where stellar magnetic fields come from, and how they 
shape the birth and life of stars and their planets.  


\section*{Acknowledgements}

This paper is dedicated to F.~Praderie (\textdagger 2009) who inspired most of the work 
presented here on imaging of large-scale magnetic topologies.  
We thank P.~Hennebelle, S.~Owocki, T.~Forveille, P.~Charbonneau and an anonymous referee for 
valuable comments that improved manuscript;  we also thank C.~Moutou, R.~Far\`es, S.~Marsden and P.~Petit 
for communicating results prior to publication.  This work was supported by the French ``Agence Nationale 
pour la Recherche'' (ANR) within the ``Magnetic Protostars and Planets'' (MaPP) project.  

\bibliographystyle{aa}
\bibliography{donati}

\end{document}